\documentclass[aps, prb, amssymb, amsmath, superscriptaddress, 
twocolumn] {revtex4-2}

\usepackage{chngcntr}

\usepackage{graphicx}
\usepackage{color}
\usepackage{amsmath}
\usepackage{enumitem}
\usepackage{amssymb}
\usepackage[hidelinks]{hyperref}
\usepackage{cancel}
\usepackage{ulem}
\usepackage{multirow}

\newcommand{\be}{\begin{equation}}
	\newcommand{\ee}{\end{equation}}

\newcommand{\bea}{\begin{eqnarray}}
	\newcommand{\eea}{\end{eqnarray}}

\newcommand{\vk}{{\bf k}}

\newcommand{\vq}{{\bf q}}
\newcommand{\p}{\partial}

\newcommand{\lp}{\left(}
\newcommand{\rp}{\right)}

\renewcommand{\vec}[1]{{\boldsymbol #1}}

\makeatother

\begin{document}
	\title{Effect of  the hyperbolic photon mode on a proximate material}
	
\author{Zhiyu Dong}
\affiliation{Department of Physics and Institute for Quantum Information and Matter, California Institute of Technology, Pasadena, California 91125}
\author{Patrick A. Lee}
\affiliation{Department of Physics, Massachusetts Institute of Technology, Cambridge, MA 02139}

\begin{abstract}

The hyperbolic mode (HM) refers to a polariton mode where the dielectric function is negative in some direction of propagation. Within a frequency window the light occupies a greatly expanded region in momentum space. We compute the photon density of states and find that the zero point fluctuation in the RMS electric field can reach $MV/cm$, comparable to the largest field used in pump probe experiments in the THz scale. The HM in hexagonal Boron Nitride (hBN) has been under intense study and  we consider placing  a material  directly on top of or sandwiched between hBN crystals. We consider two examples. First the material is a metal and we calculate the modification of  the quasi-particle spectral weight, Fermi velocity and pairing interaction. Next we consider a material that is close to the Mott transition. We find that a substantial shift in the metal-insulator transition is possible, but the effect decays rapidly with distance, so that only a few monolayers are affected. We provide estimates and suggestions for a number of materials of interest.

\end{abstract} 

\maketitle

\section{Introduction}
In a recent paper, Keren et al.\cite{keren2026cavity} reported that by placing a thin slab of hBN (with thickness of 25-100 nm) on top of an organic superconductor (SC) the superconductivity is suppressed in a region just under the hBN, as detected by a reduction of the Meissner screening current. Their modeling suggests that the SC must be modified over a substantial distance of order 400 nm or more in the top layers of the SC beneath the hBN. Inspired by this work, we study the  enhanced  zero point  fluctuation of the photon just outside the hBN, due to the presence of hyperbolic mode (HM) in hBN within a certain frequency window and consider whether this zero point fluctuation can affect physical properties of a nearby material. The study of sub-wavelength confinement of light and its strong coupling to matter has a long history. See recent reviews \cite{basov2025polaritonic, bretscher2026fluctuation} and references therein. Here we consider two examples. First, the effect on the quasi-particle spectral weight, Fermi velocity and pairing interaction in a metal and second, the  effect on a system near the  the metal insulator Mott transition. We also discuss some subtleties concerning the choice of gauge and conclude that the Coulomb gauge is the more convenient gauge choice, due to the almost longitudinal nature of the HM.

We begin by discussing the type 2 HM, where the dielectric function becomes negative for light propagating along $z$, the direction perpendicular to the plane, while it remains positive for in-plane propagation.  In hBN a polar phonon couples to light to form a polariton with transverse and longitudinal frequencies $\omega_T$ and $\omega_L$ and the dielectric function  $\epsilon(\omega)$ for propagation along z takes the form 
\be \label{dielectric def}
\tilde{\epsilon}(\omega)=\epsilon(\omega)/\epsilon_{\infty}=(\omega_L^2-\omega^2)/(\omega_T^2-\omega^2).
\ee
We define $\eta^2=\omega_L^2-\omega_T^2.$ In hBN, $\omega_T=1370 \,cm^{-1}$, $\omega_L=1611                      \,cm^{-1}$ , $\eta/\omega_T=0.61$ and $\epsilon_\infty \approx 4.5$. The mode frequency  is given by
\be\label{dispersion}
\omega_{\vec{q}\kappa}^2/\tilde{c} ^2= \vec{q}^2 + \kappa^2/\tilde{\epsilon}(\omega_{\vec{q}\kappa}),
\ee
The dispersion depends on both $\vec{q}$ and $\kappa$ 
where $\vec{q}$ is an in-plane wave vector, $\kappa$ is an out of plane vector   . For simplicity we assume the material dielectric constant $\epsilon_{\infty}$ to be the same in hBN and the organic compound and define $\tilde{c}^2=c^2/(\epsilon_{\infty}/\epsilon_0)$.

\begin{figure}
    \centering
    \includegraphics[width=1.2\linewidth]{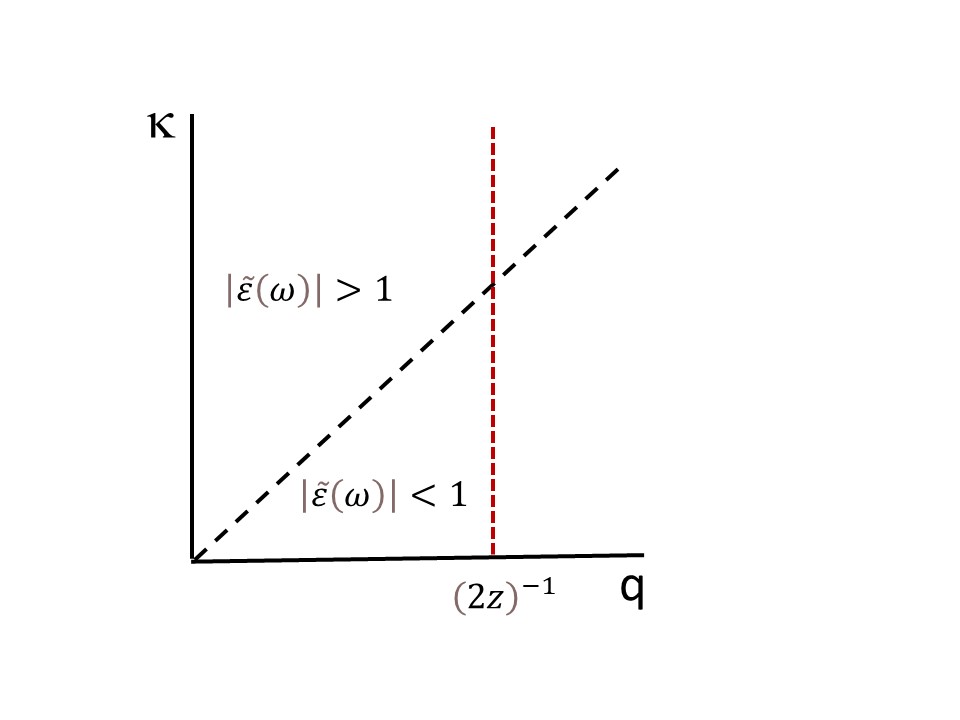}
    \caption{The  momentum phase space covered by the HM. The wave vector $\kappa$ is perpendicular to the surface and $q$ is the magnitude of the in-plane wave vector. For a given mode frequency $\omega_{\vec{q}\kappa}$ its contour is given very accurately by Eq.\ref{q vs kappa} which takes the form of a straight line through the origin with slope $\sqrt{|\tilde{\epsilon}(\omega_{\vec{q}\kappa})}|$. The dashed line separate the regions where $|\tilde{\epsilon}|$ is small or large compared with unity, corresponding to regions near $\omega_L$  and $\omega_T$ respectively. For type 2 HM, the $|\tilde{\epsilon}|>1$ region is above the dashed line, as shown in this figure, while it is below for type 1. The vertical dashed line marks the cut-off due to the exponential decay of the HM wavefunction at a distance $z$ from the hBN surface. Note that  $\vec{q}$ is a 2D vector and the HM fills the entire 3D space inside a cylinder with radius $q=1/2z$. In contrast, for ordinary photons, the constant frequency contours are circles  with radii that are typically 3 orders of magnitude smaller and invisible in this plot.}
    \label{fig:fig.1.jpg}
\end{figure}

The HM forms in the frequency range  $\omega_T< \omega_{\vec{q}\kappa} < \omega_L$ where $\tilde{\epsilon}$ is negative. We see from Eq.\ref{dispersion} that for a given $\omega_{\vec{q}\kappa}$, $\vec{q}^2$ can be much larger than the usual wave vector because it can get canceled by the second term. In fact, the left hand side is the square of the wave vector in free space for frequency between $\omega_T$ and $\omega_L$ which corresponds to wavelengths of order microns. This is 3 orders of magnitude compared with the nm scale that is typical in solid state materials. Therefore in most of the phase space of interest in this work, we can safely neglect the LHS, resulting in the relation which we will use in the rest of the paper.
\be \label{q vs kappa}
 \vec{q}^2 \approx  \kappa^2/|\tilde{\epsilon}(\omega_{\vec{q}\kappa})|,
\ee
We see that for a given mode frequency $\omega_{\vec{q}\kappa}$ the fixed frequency contours are straight lines in the 2D $q,\kappa$ plane where $q=|\vec{q}|$, as shown in Fig.\ref{fig:fig.1.jpg}. The slopes are given by $\sqrt{|\tilde{\epsilon}}|$. In contrast, outside the HM frequency window, the constant frequency contours are circles centered at the origin with a radius which is invisible in this plot, being 3 orders of magnitude smaller. (The deviation of Eq.\ref{q vs kappa}  from Eq.\ref{dispersion} occurs only on this scale: the constant frequency contour intersects the q axis as a square root, leading to  the type 2 hyperbolic contour in the 3D ($\kappa,\vq$) space.) In this way the hyperbolic nature of the dispersion  greatly enhances the momentum space corresponding to a given photon frequency within the  window between $\omega_T$ and $\omega_L$. The enhancement can be as large as $10^9$ in three dimensional space.

A similar discussion pertains to the the type 1 HM except that now it is the coefficient of the $\vec{q}^2$ term that becomes negative. We add a subscript "1" to denote this case and Eq. \eqref{dispersion} becomes $\omega_{1\vec{q}\kappa}^2/\tilde{c} ^2= \kappa^2 + \vec{q}^2/\tilde{\epsilon}_1(\omega_{\vec{q}\kappa})$ where $\tilde{\epsilon}_1(\omega)=\epsilon_1(\omega)/\epsilon_{\infty}=(\omega_{1L}^2-\omega^2)/(\omega_{1T}^2-\omega^2)$. We define $\eta_1^2=\omega_{1L}^2-\omega_{1T}^2$. In hBN this mode occurs at a lower frequency $\omega_{1T}=777 \, cm^{-1}$ and $\eta_1/\omega_{1T}=0.37$. An important difference is that instead of Eq.\ref{q vs kappa}, we have $\vec{q}^2 \approx  \kappa^2|\tilde{\epsilon}_1(\omega_{\vec{q}\kappa})|$. Now the region $|\tilde{\epsilon}_1(\omega)|>1$ lies below the dashed line in the $\kappa,q$ plane shown in Fig.\ref{fig:fig.1.jpg}. Apart from this, we shall find that the type 1 and type 2 HM contribute in very similar ways to physical phenomena.

The possible effect of the HM on electronic systems have been proposed by Ashida et al.\cite{ashida2021cavity}. Keren et al.\cite{keren2026cavity} provide a number of theoretical treatments in their paper. They computed numerically the photon local density of states at a distance z above the hBN surface and showed that an enhancement of order $10^5$ is possible, even though the enhancement decreases rapidly with increasing $z$. Instead of direct coupling to the electron, they propose a coupling via a phonon mode in the SC. In this organic SC, there is a C-C stretching phonon mode at a frequency $\omega_{ph}$ that couples strongly to the electrons which accidentally lies very close to $\omega_T$ . They find an avoided crossing of this mode to the HM and propose changes in the electronic properties, such as the electron phonon coupling.

\begin{figure}
    \centering
    \includegraphics[width=1.\linewidth]{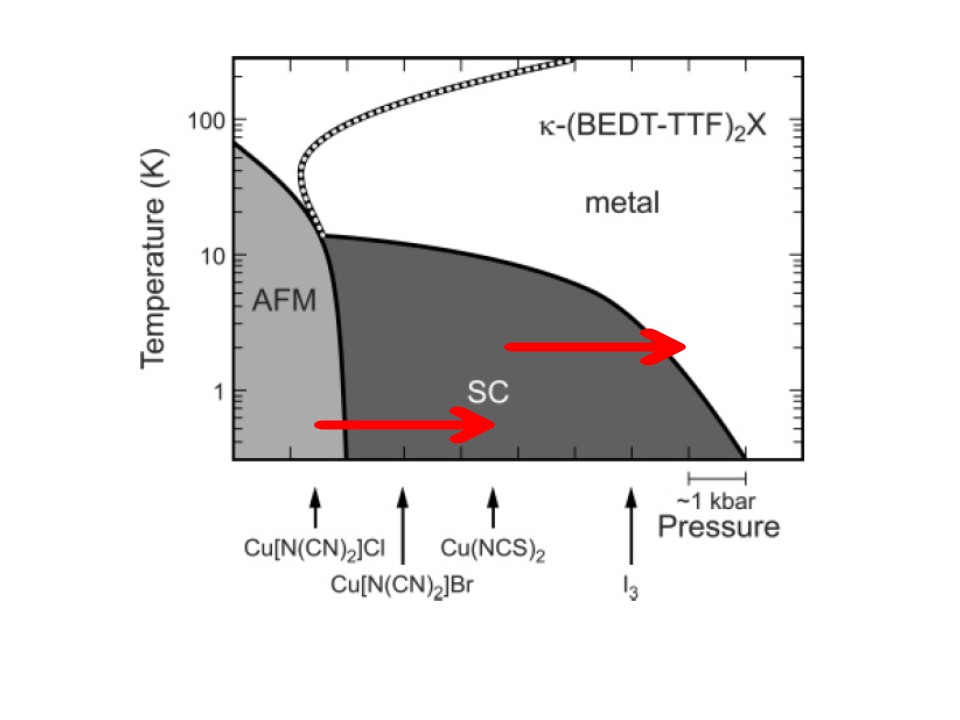}
    \caption{The generalized phase diagram for several $\kappa-$ET organic compounds. Under small pressure, the system evolves from an antiferromagentic Mott insulator to a SC before becoming a metal. Various compound are mapped  to the equivalent pressure points. In our theory the coupling to HM may push the systems along the horizontal axis towards the metallic side. The trend is marked by the red arrows. The effect decreases as the inverse of the distance to the hBN surface. Figure adapted from ref.\cite{dressel2004optical}}
    \label{fig:phase diagram.jpg}
\end{figure}

In this paper we investigate further the HM in hBN and its possible effects on an electronic system nearby. The aim is to obtain  analytic results that can lead to some understand the order of magnitude of various possible effects. 
We consider two situations. (1) The system is a Fermi liquid metal and (2) the electronic system is close to a Mott insulator. The second case is directly inspired by ref.~\cite{keren2026cavity}. It is generally accepted that  the organic SC studied there  is close to the Mott transition, as shown in the generalized phase diagram shown in  Fig.\ref{fig:phase diagram.jpg}. The Mott transition is commonly described by the Hubbard model at half filling where an on site $U$ term is added to the hopping $t$ term on a lattice. We use a relatively simple and transparent description of the half-filled  Hubbard model due to Florence and Georges \cite{florens2004slave} and  show that coupling of the charge degrees of freedom in this model to the HM can lead to a considerable change  of the correlation, and may push the system towards the more metallic  side. The effect can be quite strong for  the first few layers, so that for the right material, a significant change in physical properties is possible for the first few layers of a layered electronic system that is in close proximity to the hBN. Unfortunately the effect decays exponentially with distance,  making it challenging to explain the experiment of keren et al.\cite{keren2026cavity}

While  we are not claiming to understand this particular experiment, we believe other experimental applications may be possible, as will be discussed in the discussion section. 
Note that our mechanism directly couples the HM to the electronic system and does not require the accidental resonance with a phonon mode. In fact we argue that the phonon mechanism proposed by Keren et al. \cite{keren2026cavity} is unlikely to be important. They considered the effect of the particular  phonon on the electronic system, which is of course substantial, but  the right question to ask in this particular discussion is what is the effect of the hybridization with the HM on the coupling of this phonon to the electronic system. The answer to the question is that the relative change is very small, less the $(\Delta_1/\omega_{ph})^2$ where $\Delta_1$ is the anti-crossing gap. This is probably of order $10^{-5}$ and negligible.

\section{The HM photon local density of states (DOS) }\
We begin with  the operator for  the projection of the electric field  of the  HM mode to the plane for propagation with a background $\epsilon_{\infty}$. 
\begin{align}
\boldsymbol{E}_{\parallel}(\boldsymbol{r})=&\frac{i}{\sqrt d}\sum_{\kappa>0}\int\!\frac{d^2q}{(2\pi)^2}
\hat{\mathbf{q}}\sqrt{\frac{\omega_{\mathbf{q}\kappa}}{2\epsilon_\infty}}\,
 f(\mathbf{q},\kappa,z)
\nonumber\\
&\times\left(a_{\mathbf{q}\kappa}e^{i\mathbf{q}\cdot\boldsymbol{r}}
-a^\dagger_{\mathbf{q}\kappa}e^{-i\mathbf{q}\cdot\boldsymbol{r}}\right).
\label{eq:Eparallel-operator}
\end{align}
where $f(\vq,\kappa,z)$ is the mode function. Note that the mode is polarized along the propagation direction in the plane. \cite{ashida2021cavity}

We write the photon density of states (DOS) at a position $z$ outside the hBN slab as
\be\label{DOS def}
\rho(\omega, z)= \int_{0}^{\Lambda_q} \frac{d\kappa}{2\pi}\int_{0}^{\Lambda_q} \frac{d\vec{q}}{(2\pi)^2}\,\delta(\omega-\omega_{\vec{q}\kappa})  |f(\vec{q,\kappa,z)}|^2
\ee
 We assume the thickness $d>>z$ and take the continuum limit for the $\kappa$ integration to replace the sum over mode indices. The upper limit $\Lambda_q$ is the inverse of a lattice scale below which the dielectric function approximation is valid. 
 For $z$ located outside the sample surface the local wavefunction is given by (see Appendix \ref{appendix: normalization} for details.)
 \be \label{wf main}
|f(\vec{q,\kappa,z)}|^2=f_{\vq\kappa}^2cos^2(\kappa z) e^{-2 \nu(q)z} \approx f_{\vq\kappa}^2 e^{-2 \nu(q)z}
\ee 
The second step is justified in most cases of interest, where  $1/z$ serves as the upper cut-off for the $\kappa$ integration.
The normalization factor $f_{\vq\kappa}^2$ is given in Appendix \ref{appendix: normalization} and 
\be \label{nu def}
 \nu_{\vec{q}}=\sqrt{q^2-\omega_{\vec{q}\kappa}^2/\tilde{c}^2} \approx q.
\ee
The last approximation is justified due to the large value of $\tilde{c}$, as discussed earlier.

First we consider $\omega_L-\omega$ small and introduce the small difference
\be\label{Delta omega}
\Delta_{\omega}^2 = \omega_L^2-\omega^2.
\ee
in which case  $|\tilde{\epsilon}|<<1$, and $\tilde{\epsilon}$ can be expanded as
\be\label{eps expansion}
\tilde{\epsilon}=-\Delta_{\omega}^2/\eta^2
\ee
which vanishes linearly for $\omega$ just less than $\omega_L$ . We write
\be \label{q vs kappa 1}
 \frac{\vec{q}^2}{\kappa^2} \approx \frac{1}{|\tilde{\epsilon}(\omega_{\vec{q}\kappa})|}\approx \frac{\eta^2}{\Delta_{\vq}^2},
\ee
where we define 
\be\label{delta q}
\Delta_\vec{q}^2=\omega_L^2-\omega_{\vec{q}\kappa}^2 
\ee
which is similar to Eq. \eqref{Delta omega} but with $\omega$ evaluated at $\omega_{\vec{q}\kappa}$ so that it is  a function of $q$ for fixed $\kappa$. 
In Appendix \ref{appendix: normalization} we show that in this regime $f_\vec{q,\kappa}^2/2$ is given approximately by $\eta^2/\omega_L^2$. Then Eq. \eqref{DOS def} becomes
\be\label{DOS int1}
\rho(\omega)= \frac{\eta^2}{\omega_L^2}\int_{0}^{\Lambda} \frac{d\kappa}{\pi}\int_{0}^{\Lambda} \frac{d(q^2)}{2\pi}\,\delta(\Delta_{\omega}^2-\Delta_\vec{q}^2) \omega  e^{-2qz}
\ee
By changing the integration variable from $q^2$ to $\Delta_\vec{q}^2$  using the second part of Eq.\ref{q vs kappa 1}, the integral is evaluated. (See appendix \ref{appendix:DOS} for details.) For frequency a little less than  $\omega_L$, it diverges as
\be \label{DOS1Sum}
\rho(\omega,z)= \frac{2\eta \omega }{2\pi^2 \omega_L^2\sqrt{\omega_L^2-\omega^2}}\frac{1}{(2z)^3}, 
\ee
A similar treatment can be made for frequency a little larger than $\omega_T$. The normalization factor $f_\vec{q,\kappa}^2/2$ is given in Appendix \ref{appendix: normalization} Eq.\ref{f near T}.  The DOS now vanishes as
\be \label{DOST1sum}
\rho(\omega, z)= \frac{2\omega }{2\pi^2 \omega_T^2 }\sqrt{\frac{\omega^2-\omega_T^2}{\eta^2}}\frac{1}{(2z)^3}.
\ee
By integrating these equations over $\omega$ we estimate the equal time mean square electric field at position $z$
\be \label{|EE|}
\langle |E(z)|^2 \rangle =\frac{1}{\epsilon_\infty}\int d\omega \,\omega\rho(\omega)\\
\approx \frac{1}{4\pi \epsilon_\infty }\frac{\eta^2}{2\pi \sqrt{2}\omega_L}\frac{1}{z^3}.
\ee
We perform the integration approximately by first using the form Eq.\ref{DOS1Sum} which has an integrable divergence near $\omega_L$ and cutoff the integral at $\omega_L^2-\omega^2=\eta^2/2$. Integration over the form Eq.\ref{DOST1sum} gives the same functional form but with a smaller prefactor. The $\langle |E(z)|^2\rangle$ can be written in the suggestive form $(\eta/ez) (e/(4\pi \epsilon_{\infty}z^2)\frac{\eta}{2\pi \sqrt{2}\omega_L}$ so the RMS value can be read off  as volt/cm. Setting $z=0.5\rm{nm}$ which is a typical layer spacing, and using $\eta \approx 0.1 \rm{eV}$ we estimate an RMS value of $10^{6}\, \rm{V/cm}$. This is a very large value, comparable  with the largest that can be generated in the Lab at THz scale. On the other hand, the HM is mainly longitudinal, so that the electric field comes from the Coulomb potential created by the zero point motion of the ions. From this point of view the large electric field is not unreasonable. We must also keep in mind that this is vacuum fluctuation which is part of the ground state, and cannot create absorption or dissipation. It can contribute to virtual process and its large value gives us hope that direct modification of electronic systems is possible, if $z$ is of order 0.5 nm. On the other hand, the $z^{-3}$ suppression of the mean square electric field greatly limit the range of this effect.

\section{Effective interaction between electrons in a metal due to coupling to HM.}
\label{sec:Coulomb gauge}


We consider metal coupling to HM. The metal is treated as a two-dimensional Fermi liquid with
\begin{equation}
H_e=\sum_{\mathbf{k}}\xi_{\mathbf{k}}c^\dagger_{\mathbf{k}}c_{\mathbf{k}},
\qquad
\rho_{\mathbf{q}}=\sum_{\mathbf{k}}{\cal F}(\mathbf{k},\mathbf{q})c^\dagger_{\mathbf{k}+\mathbf{q}}c_{\mathbf{k}}.
\label{eq:electron-hamiltonian}
\end{equation}
The form factor ${\cal F}(\mathbf{k},\mathbf{q})=\langle u_{\mathbf{k}+\mathbf{q}}|u_{\mathbf{k}}\rangle$ will be kept in the first few equations and then set to one, as in the original single-band calculation.

In the Coulomb gauge,
\begin{equation}
\boldsymbol{\nabla}\cdot\boldsymbol{A}=0,
\qquad
\boldsymbol{E}=-\boldsymbol{\nabla}\phi-\partial_t\boldsymbol{A},
\label{eq:coulomb-gauge}
\end{equation}
so the longitudinal electric field is carried by the scalar potential, whereas the vector potential contains only the transverse component.  The two couplings are
\begin{equation}
H_{\rm int}=-e\int d^2r\,\rho\phi
+e\int d^2r\,\boldsymbol{j}\cdot\boldsymbol{A}_{\perp}.
\label{eq:two-couplings}
\end{equation}

As the HM is in the quasi-static regime ($\omega\ll \tilde{c}q$), it couples to electrons mainly through the scalar potential's $\rho \phi$ term because the electric field is predominantly longitudinal even before projection to the plane. Consider a mode propagation along $\boldsymbol{Q}=(q,0,\kappa)$. From Maxwell equation $\boldsymbol{Q}\times(\boldsymbol{Q}\times\boldsymbol{E})
+\frac{\omega^2}{\widetilde c^{\,2}}\,
\widetilde{\boldsymbol\epsilon}(\omega)\boldsymbol{E}=0$, we see that 
\begin{equation}
\left|\frac{E_{\perp}}{E_{\parallel}}\right|
\sim
\frac{\omega^2}{\widetilde c^{\,2}}
\frac{\left|q\kappa
(\widetilde\epsilon_z-\widetilde\epsilon_t)\right|}
{(q^2+\kappa^2)^2}\ll1.
\label{eq:Etransverse-suppression}
\end{equation}
where $E_{\perp}$ and $E_{\parallel}$ are the field components perpendicular and parallel to $\boldsymbol{Q}$. This can also be seen directly from $\boldsymbol{\nabla}\cdot\boldsymbol{}D=qE_x-\kappa\tilde{\epsilon}E_z=0$ and using Eq.~\ref{q vs kappa} to show that $\boldsymbol{\nabla}\times\boldsymbol{E}=0$. The correction is proportional to $1/\tilde{c}^2$ and negligible.

A one-HM self-energy or exchange diagram contains two such vertices, so the transverse-vector-potential contribution is suppressed by the square of Eq.~\eqref{eq:Etransverse-suppression}.  We therefore retain $-e\rho\phi$ and drop the transverse current coupling.  The $\boldsymbol{A}_{\perp}^{\,2}$ term does not enter the one-HM exchange considered below.

The scalar potential is obtained from the frequency-dependent Poisson problem
\begin{equation}
-\partial_\alpha\!\left[\epsilon_{\alpha\beta}(z,\omega)\partial_\beta\phi_{\mathbf{q}}(z,\omega)\right]
=\rho^{\rm ext}_{\mathbf{q}}(z,\omega),
\label{eq:poisson}
\end{equation}
with the boundary conditions of the hBN--metal geometry. The mode structure is solved by Ashida et al.\cite{ashida2021cavity}.  Its poles are the hyperbolic polariton modes.  The  in-plane longitudinal electric field of the normalized HM modes is given by Eq.
\ref{eq:Eparallel-operator}. 
In the Coulomb gauge this longitudinal field is carried by the scalar potential,
\begin{equation}
\boldsymbol{E}_{\parallel}=-\boldsymbol{\nabla}_{\parallel}\phi,
\qquad
\phi_{\mathbf{q}\kappa}
=-\frac{1}{q}\sqrt{\frac{\omega_{\mathbf{q}\kappa}}{2\epsilon_\infty}}\,
 f(\mathbf{q},\kappa,z),
\label{eq:phi-amplitude}
\end{equation}
so the scalar-potential operator at the metallic plane is
\begin{align}
\phi(\boldsymbol{r})=&-\frac{1}{\sqrt d}\sum_{\kappa>0}\int\!\frac{d^2q}{(2\pi)^2}
\frac{1}{q}\sqrt{\frac{\omega_{\mathbf{q}\kappa}}{2\epsilon_\infty}}\,
 f(\mathbf{q},\kappa,z)
\nonumber\\
&\times\left(a_{\mathbf{q}\kappa}e^{i\mathbf{q}\cdot\boldsymbol{r}}
+a^\dagger_{\mathbf{q}\kappa}e^{-i\mathbf{q}\cdot\boldsymbol{r}}\right).
\label{eq:phi-operator}
\end{align}
The normalization length $d$ is chosen so that a closed mode sum is $(1/d)\sum_\kappa\to\int d\kappa/(2\pi)$.  Equation~\eqref{eq:phi-operator} is the pole expansion of the frequency-dependent scalar Green function generated by the dielectric polarization. Our approach is equivalent to that used by Riolo et al.~\cite{riolo2025tuning} who focus on the pole structures in the solution of Eq.~\ref{eq:poisson}. They employ the RPA approximation for the screening  and presented detailed numerical solutions. We use the simpler  static screening approximation in our treatment below  and our mainly analytic solutions complement their results.~\cite{riolo2025tuning}~\cite{riolo2025polariton}

The density coupling is therefore
\begin{align}
H_{e\phi}&=-e\int d^2r\,\rho(\boldsymbol{r})\phi(\boldsymbol{r})
\nonumber\\
&=\frac{1}{\sqrt d}\sum_{\mathbf{q},\kappa>0}g_{\mathbf{q}\kappa}\,
\rho_{-\mathbf{q}}\left(a_{\mathbf{q}\kappa}+a^\dagger_{-\mathbf{q}\kappa}\right),
\label{eq:rho-phi-hamiltonian}
\end{align}
with
\begin{equation}
g_{\mathbf{q}\kappa}
=\frac{e}{q}\sqrt{\frac{\omega_{\mathbf{q}\kappa}}{2\epsilon_\infty}}\,
 f(\mathbf{q},\kappa,z).
\label{eq:coulomb-vertex}
\end{equation}
The vertex contains $\omega_{\mathbf{q}\kappa}/q$. Apart from the Bloch overlap ${\cal F}(\mathbf{k},\mathbf{q})$, it is independent of the electronic energies. As an aside, we point out  that had we used the temporal (Weyl) gauge, the $\vec A \cdot \vec j$ vertex would replace  $\omega_{\mathbf{q}\kappa}/q$ by $(\xi_{\mathbf{k}+\mathbf{q}}-\xi_{\mathbf{k}})/q$ which can lead to UV divergence. The issue of gauge choice will be discussed in a later section, but here we concentrate on the Coulomb gauge.

Integrating out the HM oscillators gives the single HM-mediated density-density interaction,
\begin{align}
V(\mathbf{q},i\Omega_n)
&=-\frac{1}{d}\sum_{\kappa>0}
\frac{2\omega_{\mathbf{q}\kappa}|g_{\mathbf{q}\kappa}|^2}
{\Omega_n^2+\omega_{\mathbf{q}\kappa}^2}
\label{eq:V-exact}
\end{align}
In the rest of this section we consider the case where the metal is surrounded by the HM material on both sides. From the mode profile we find,
\begin{equation}
|f(\mathbf{q},\kappa,z)|^2
=f_{\mathbf{q}\kappa}^2\frac{\cos^2(\kappa z)}{\cosh^2(qz)},
\label{eq:mode-profile}
\end{equation}
the normalization integral used below is given in Eq.~\ref{I kappa 1}
\begin{equation}
\int_0^\infty\frac{d\kappa}{2\pi}
 f_{\mathbf{q}\kappa}^2
\simeq
\frac{q}{2\sqrt2}\frac{\eta^2}{\omega_T\omega_L}
\sqrt{\frac{\omega_T}{\omega_L}}.
\label{eq:kappa-integral}
\end{equation}
For the two-sided profile in Eq.~\eqref{eq:mode-profile}, the factor $\cos^2(\kappa z)$ is replaced by unity within the same mode-profile approximation. The effect for the one-sided geometry will be half as large.
Contributions from different hyperbolic windows add in Eq.~\eqref{eq:V-exact}.

Since $\omega_{\mathbf{q}\kappa}$ lies in a finite range between $\omega_T$ and $\omega_L$ we can replace it by the average value $\bar{\omega}$ and perform the $\kappa$ integration in Eq. ~\eqref{eq:V-exact}. Using Eqs. ~\eqref{eq:mode-profile},~\eqref{eq:kappa-integral} we obtain 
\be  
\label{eq:V q only}
V(\mathbf{q},i\Omega_n)=
-\frac{e^2}{q \epsilon_\infty}\frac{\bar{\omega}^2}{\Omega_n^2 +\bar{\omega}^2}\left[\frac{\eta^2}{2\sqrt2\omega_T\omega_L}
\sqrt{\frac{\omega_T}{\omega_L}}\frac{1}{\cosh^2(qz)}\right]
\ee
It is interesting to compare this with the Fr\"ohlich effective interaction between electrons in a two dimensional metal that is surrounded by exchanging an longitudinal optical phonon $\omega_L$ in an  polar dielectric
\begin{align}
\label{eq:Frohlich}
V_F(\mathbf{q},i\Omega_n)&=\frac{e^2}{2q \epsilon_\infty}(\frac{1}{\epsilon(\omega)}-\frac{1}{\epsilon_\infty})
\nonumber\\
&=-\frac{e^2}{2q \epsilon_\infty}\frac{\omega_L^2}{\Omega_n^2 +\omega_L^2}\left[\frac{\eta^2}{\omega_L^2}\right]
\end{align}
In the second line we have used the frequency dependent dielectric function shown in Eq.~\eqref{dielectric def}, assuming an isotropic medium. Apart from the terms in [ ], which are comparable in magnitude for $z=0$, the HM form given in Eq.~\ref{eq:V q only} takes the same form as the Fr\"ohlich interaction. Therefore we expect their effect to be similar in order of magnitude. We will carry out the detailed evaluation of the HM case for finite $z$ below.

In the Fr\"ohlich interaction for a metal we should screen the bare interaction by the metallic screening. In the static approximation which is valid for $\omega_L \ll E_F$ we replace $1/q$ by $1/(q+\kappa_{TF})$
 where $\kappa_{TF}=\frac{e^2}{4 \pi \epsilon_\infty} 2\nu_0$ and $\nu_0$ is the 2D density of states. We will do the same for the HM case in Eq.~\ref{eq:V q only}.

We define the constant
\be \label{C_v}
C_v=\frac{e^2}{4\pi\epsilon_\infty v}
\frac{1}{2\sqrt2}\frac{\eta^2}{\omega_T\omega_L}
\sqrt{\frac{\omega_T}{\omega_L}},
\ee
In the rest of this section we define for fermions $C=C_{v_F}$.
and the HM-mediated interaction can be simplified to
\be\label{eq:V q only TF}
V(\mathbf{q},0)=-\frac{4\pi v_FC}{(q+\kappa_{TF})\cosh^2(qz)}.
\ee
The self energy generated by this interaction is the exchange
diagram
\begin{equation}
\Sigma(\mathbf{k},i\omega_n)=-T\sum_{\Omega_n}
\int\!\frac{d^2q}{(2\pi)^2}\,
\frac{V(\mathbf{q},i\Omega_n)}
{i\omega_n+i\Omega_n-\xi_{\mathbf{k}+\mathbf{q}}} .
\label{eq:Sigma-matsubara}
\end{equation}
Carrying out the frequency sum, the poles of $V$ at
$i\Omega_n=\pm\bar\omega$ have residues $\mp\bar\omega V(\mathbf{q},0)/2$. After analytic continuation 
$i\omega_n\to\omega+i0^+$ we find,
\begin{align}
\Sigma^R(\mathbf{k},\omega)=&
-\frac{\bar\omega}{2}\int\!\frac{d^2q}{(2\pi)^2}
V(\mathbf{q},0)
\nonumber\\
&\times\bigg[
\frac{1-n_F(\xi_{\mathbf{k}+\mathbf{q}})}{\omega-\xi_{\mathbf{k}+\mathbf{q}}-\bar\omega+i0^+}
\nonumber\\
&+
\frac{n_F(\xi_{\mathbf{k}+\mathbf{q}})}{\omega-\xi_{\mathbf{k}+\mathbf{q}}+\bar\omega+i0^+}
\bigg].
\label{eq:retarded-self-energy}
\end{align}
An additive constant is absorbed into the chemical potential. The residue is controlled by the frequency derivative of the self
energy on the Fermi surface,
\begin{equation}
\lambda_0=-\left.\partial_\omega\operatorname{Re}\Sigma^R\right|
_{\mathbf{k}=\mathbf{k}_F,\,\omega=0} .
\label{eq:lambda-def}
\end{equation}
Differentiating Eq.~\eqref{eq:retarded-self-energy}, the particle
and the hole term combine into
\begin{equation}
\lambda_0=-\frac{\bar\omega}{2}\int\!\frac{d^2q}{(2\pi)^2}\,
\frac{V(\mathbf{q},0)}
{\left(|\xi_{\mathbf{k}_F+\mathbf{q}}|+\bar\omega\right)^2} ,
\label{eq:lambda-q}
\end{equation}
which is positive-valued as $V(\vq,0)$ is negative.
Inserting Eq.~\eqref{eq:V q only TF}, with $x=qz$ and
$\xi_{\mathbf{k}_F+\mathbf{q}}\simeq v_Fq\cos\theta$ where $\theta$ is the angle between $\mathbf{q}$ and $\mathbf{k_F}$. In this paper we will be interested in the regime   $k_Fz\gg1$. Since $q$ is limited by $1/2z$ by the $\cosh^2(qz)$ factor, this means that only a thin shell in the momentum space near the Fermi surface is accessible.
\begin{align}
\lambda_0=&\,C\alpha
\int_0^\infty\frac{dx}{\cosh^2x}\,\frac{x}{x+K}
\int_0^{2\pi}
\frac{\frac{d\theta}{2\pi}}{\left(1+\alpha x|\cos\theta|\right)^2} .
\label{eq:lambda-dimensionless}
\end{align}
where we have introduced an important dimensionless parameter that characterizes the system
\be
\alpha=\frac{v_F}{z\bar\omega}
\ee
and a second parameter
$K=\kappa_{TF}z$, which compares the distance to
the slab with the screening length .
The mode's exponential profile ($1/\cosh^2x$ term) provides an upper limit to the $x$ integral at $x\simeq1$, 
and the angular integral in Eq. \eqref{eq:lambda-dimensionless} can be replaced by 
unity for $\alpha x\ll1$ and $2/\pi\alpha x$ for $\alpha x\gg1$. With these simplifications Eq.  ~\ref{eq:lambda-dimensionless} can easily be evaluated in 4 different regimes:
\begin{equation}
\lambda_0\simeq
\begin{cases}
C\alpha , & \alpha\ll1,\ K\ll1,\\
\frac{C\alpha\ln2}{K} , & \alpha\ll1,\ K\gg1,\\
\frac{2C}{\pi}\ln\dfrac{1}{\max(K,\alpha^{-1})} ,
& \alpha\gg1,\ K\ll1,\\
\frac{2C}{\pi\,K} , & \alpha\gg1,\ K\gg1 ,
\end{cases}
\label{eq:lambda-asymptotics}
\end{equation}
The quasiparticle residue is given by
\begin{equation}
Z=\frac{1}{1+\lambda_0}.
\label{eq:Z-fermion}
\end{equation}

The velocity renormalization separates into the same two
regimes.  For $K\gg1$ the screening momentum lies beyond the
range $q\lesssim1/z$ allowed by the profile , so that the interaction is $q$ independent,
\begin{equation}
V(\mathbf{q},0)\simeq V_0\equiv-\frac{4\pi v_FC}{\kappa_{TF}} ,
\label{eq:V-const}
\end{equation}
and the momentum drops out of
Eq.~\eqref{eq:retarded-self-energy} altogether: shifting the momentum
variable $\vk+\vq \rightarrow \vq$ , the self energy is an integral over energy alone,
\begin{align}
\Sigma^R(\mathbf{k},\omega)=& - \frac{\bar\omega V_0\nu_0}
{2\pi k_Fz}
\int_{-\Lambda}^{\Lambda}\!d\xi
\left[\frac{\Theta(\xi)}{\omega-\xi-\bar\omega+i0^+}\right.
\nonumber\\
&\qquad\quad\left.+\frac{\Theta(-\xi)}{\omega-\xi+\bar\omega+i0^+}
\right] ,
\label{eq:Sigma-1d}
\end{align}
where $\nu_0$ is the density of states on the Fermi surface and $\Lambda\equiv v_F/z=\alpha\bar\omega$ the energy range spanned by
$q<1/z$. The factor $1/\pi k_Fz$ comes from the angle integration over the Fermi surface and captures the fraction  that is reached by the interaction. Just like the familiar case of electron-phonon coupling, the self-energy is k-independent for $K\gg 1$, and  only has  frequency
dependence. Therefore the velocity correction takes the usual form 
\begin{equation}
\frac{v_F^*}{v_F}=Z .
\label{eq:vF-reduced}
\end{equation}
For $K\ll1$, $V(\mathbf{q},0)$ varies approximately as $1/q$,
Eq.~\eqref{eq:Sigma-1d} does not apply, and we 
have to evaluate the k dependence of the self energy. Differentiating Eq.~\eqref{eq:retarded-self-energy} at $\omega=0$
with respect to $\mathbf{k}$, and using
$\hat{\mathbf{k}}_F\!\cdot\!\boldsymbol{v}_{\mathbf{k}_F+\mathbf{q}}/v_F
=1+(q/k_F)\cos\theta\simeq1$ due to the fact that we are interested in the range  $q\lesssim1/z\ll k_F$, the derivative of the
bracket is
\begin{equation}
\frac{d}{d\xi}
\left[-\frac{\Theta(\xi)}{\xi+\bar\omega}
+\frac{\Theta(-\xi)}{\bar\omega-\xi}\right]
=\frac{1}{\left(|\xi|+\bar\omega\right)^2}-\frac{2}{\bar\omega}\delta(\xi) ,
\label{eq:static-derivative}
\end{equation}
so that
\begin{equation}
\frac{d\Sigma^R(\mathbf{k},\omega=0)}{d\xi_k} \equiv F_k=\lambda_0
+\int\!\frac{d^2q}{(2\pi)^2}V(\mathbf{q},0)\,
\delta(\xi_{\mathbf{k}_F+\mathbf{q}}) ,
\label{eq:Fk-two-terms}
\end{equation}

The first term $\lambda_0$ is given in Eq.~\eqref{eq:lambda-q}.  In the
second term, we find  $\int_0^{2\pi}d\theta\,\delta(v_Fq\cos\theta)=2/v_Fq$ , the factor 2 coming from the fact that the delta function is satisfied by $\theta=\pm \pi/2$. Changing variable to $x=qz$, 
\begin{equation}
-\int\!\frac{d^2q}{(2\pi)^2}V(\mathbf{q},0)\,
\delta(\xi_{\mathbf{k}_F+\mathbf{q}})
=\frac{2C}{\pi}\int_0^\infty\frac{dx}{(x+K)\cosh^2x} .
\label{eq:Fk-delta}
\end{equation}
For $\alpha\gg1$ the angular average in
Eq.~\eqref{eq:lambda-dimensionless} is $2/\pi\alpha x$ for $x\gg\alpha^{-1}$,
where the integrand of $\lambda_0$ becomes that of Eq.~\eqref{eq:Fk-delta}.  The
two therefore cancel. However, in the regime of $x<\alpha^{-1}$ they do not cancel, so $F_k$ is finite. 

\emph{Pairing by exchanging HM:}
The HM-mediated interaction of Eq.~\eqref{eq:V-exact} 
is a
retarded attraction exactly as the Fr\"ohlich interaction is, with $\bar\omega$ in place of
$\omega_L$ as discussed in Sec.\ref{sec:Coulomb gauge}. It therefore gives rise to pairing like the conventional
phonon, with the HM frequency $\bar\omega$ as the cutoff.  

The pairing strength is the projection of Eq.~\eqref{eq:V q only TF} onto the s-wave Cooper
channel, that is its average over the angle $\phi$ between the two momenta on
the Fermi surface, $g_{0} = \nu_0\langle -V(\mathbf{q}_\phi,0)\rangle_\phi$ with
$q_\phi=2k_F\sin(\phi/2)$ and $\nu_0=k_F/2\pi v_F$ the density of states per spin. In fact, this $g_0$ is nothing  but the second term in Eq.\eqref{eq:Fk-two-terms} because by changing the integration variable from  $\mathbf{q}$ to $\mathbf{k_F}+\mathbf{q}$, we find $-\int\!\frac{d^2q}{(2\pi)^2}V(\mathbf{q},0)\,
\delta(\xi_{\mathbf{k}_F+\mathbf{q}}) =-\nu_0 \int\!\frac{d\phi}{2\pi}d\xi' V(\mathbf{q}_\phi,0)\,
\delta(\xi') =\nu_0 \langle -V(\mathbf{q}_\phi,0)\rangle_\phi=g_0 $ . So we can re-express the $F_k$ and velocity correction using $g_0$
\begin{equation}
F_k=\lambda_0-g_0 ,
\qquad
\frac{v_F^*}{v_F}=\frac{1+\lambda_0-g_0}{1+\lambda_0} 
\label{eq:Fk-is-lambda-minus-g}
\end{equation}
Below we calculate $g_0$  directly from Eq.\eqref{eq:Fk-delta}:
\begin{align} 
g_{0}&= \frac{2C}{\pi}\int_0^\infty\frac{dx}{(x+K)\cosh^2x}
\simeq \frac{2C}{\pi}\ln\!\left(1+\frac{1}{K}\right)
\nonumber\\
&\simeq\frac{2C}{\pi}
\begin{cases}
\ln\dfrac{1}{K} , & K\ll1 ,\\[3mm]
\dfrac{1}{K} , & K\gg1 .
\end{cases}
\label{eq:gsc}
\end{align}
Comparing with Eq.~\eqref{eq:lambda-asymptotics}, the two couplings coincide, $g_0\simeq\lambda_0$, only when $\alpha\gg\max(1,K^{-1})$, and $g_0>\lambda_0$ otherwise. 
The reason is that the two integrands agree only where the angular average of Eq.~\eqref{eq:lambda-dimensionless} has reached its large argument form $2/\pi\alpha x$. This is achieved when $\alpha x\gg1$ over the range that dominates the integral in Eq.~\eqref{eq:lambda-dimensionless}, which is $x\simeq1$ for $K\gg1$ and $K<x<1$ for $K\ll1$.  Hence $\alpha\gg1$ is needed in both cases, and $\alpha K\gg1$ in addition when $K\ll1$. 


We note that, as the kernel is
forward peaked, the couplings in the finite-angular-momentum channels,
$g_\ell=\nu_0\langle -V\cos\ell\theta\rangle$, are nearly degenerate,
$g_\ell\simeq g_{0}$ for $\ell\lesssim k_Fz$, and split only at
order $(\ell/k_Fz)^2$.  

With the cutoff $\bar\omega$ and the mass renormalization $Z^{-1}=1+\lambda_0$
of Eq.~\eqref{eq:Z-fermion},
\begin{equation}
T_c\simeq\bar\omega\,
\exp\!\left[-\frac{1+\lambda_0}{g_{0}-\mu^*}\right] ,
\qquad
\mu^*=\frac{\mu}{1+\mu\ln(E_F/\bar\omega)} .
\label{eq:Tc}
\end{equation}
We now estimate these quantities in realistic systems. For the metal we take parameters close to those of NbSe$_2$
~\cite{johannes2006fermi,calandra2009effect}, $v_F\simeq1.5\times10^5$~m/s, that is $v_F\simeq0.1$~eV\,nm, and $k_F\simeq5$~nm$^{-1}$, so that $E_F=v_Fk_F/2\simeq0.25$~eV, and with $\nu_0=k_F/2\pi v_F$ the screening momentum is $\kappa_{TF}=(e^2/4\pi\epsilon_\infty)k_F/\pi v_F$. The slab is placed at $z=0.4$~nm, close to the van der Waals contact, so that $k_Fz\simeq2$.

For hBN, the two hyperbolic mode frequencies are $\omega_T=1370$~cm$^{-1}$, $\omega_L=1611$~cm$^{-1}$ and $\eta/\omega_T=0.61$, the numerical factor $(2\sqrt2)^{-1}(\eta^2/\omega_T\omega_L)\sqrt{\omega_T/\omega_L}$ is 0.10, and with $\epsilon_\infty=4.5$, so that $e^2/4\pi\epsilon_\infty=0.32$~eV\,nm, we write $C$ as a product of fine structure constant $e^2/4\pi\epsilon_\infty c$ and $c/v_F$ to find $C
\simeq0.32$,
and $\kappa_{TF}=(e^2/4\pi\epsilon_\infty)k_F/\pi v_F
\simeq5.1$~nm$^{-1}$. With $\bar\omega\simeq0.17$~eV, $v_F/\bar\omega\simeq0.6$~nm, so that $\alpha\simeq1.5$ and $K\simeq2.1$, giving $g_0\simeq0.08$ and $\lambda_0\simeq0.04$. Here $K$ is of order unity and $\alpha$ is not large either, we have evaluated  $\lambda_0$ using Eq.~\eqref{eq:lambda-dimensionless} instead of the limiting forms Eq.~\eqref{eq:gsc}.

For Bi$_2$Se$_3$, $\eta^2/\omega_T\omega_L=1.84$ and $\sqrt{\omega_T/\omega_L}=0.66$ give a larger numerical factor, $(2\sqrt2)^{-1}(\eta^2/\omega_T\omega_L)\sqrt{\omega_T/\omega_L}=0.43$, but $\epsilon_\infty=29$ gives $e^2/4\pi\epsilon_\infty=0.05$~eV\,nm, so that $C
\simeq0.22$ and $\kappa_{TF}=(e^2/4\pi\epsilon_\infty)k_F/\pi v_F
\simeq0.80$~nm$^{-1}$. With $\omega_L=146$~cm$^{-1}$ and $\bar\omega\simeq13$~meV, $v_F/\bar\omega\simeq7.6$~nm, so that $\alpha\simeq19$ and $K\simeq0.32$, giving $g_0\simeq0.18$ and $\lambda_0\simeq0.16$, and $Z\simeq0.86$. Here $\alpha$ is large enough but $K$ is not, and again Eq.~\eqref{eq:lambda-dimensionless} has been used.


In Table~\ref{tab:kF} we study how the result depends on carrier density. We consider a parabolic band with mass $m^*=2m_e$. 
The screening momentum $\kappa_{TF}=(e^2/4\pi\epsilon_\infty)2\nu_0=(e^2/4\pi\epsilon_\infty)m^*/\pi\hbar^2$ is independent of $k_F$, so  $K$ is fixed.  Therefore, $C\propto1/v_F\propto1/k_F$, so that $g_0$ grows as the density is reduced, reaching $0.22$ on Bi$_2$Se$_3$ at $k_F=3\,$nm$^{-1}$.  However, the growth is limited. When $k_F$ is too small, system enters the regime  $k_Fz<1$ and the scattering is no longer forward. This is outside the regime  we have assumed in this paper and the $g_0$ expression given in Eq.~\eqref{eq:gsc} is no longer valid. Instead, we should use $g_0 \sim \frac{2k_FC}{\pi}\int_0^{\pi}d\theta
\lp 2k_F \sin\theta/2+\kappa_{\rm TF}\rp^{-1 }$, which no longer increases upon decreasing $k_F$ further.
In this expression, when $k_F<\kappa_{TF}$, $g_0$ saturates as $g_0\sim \frac{2Ck_F}{\kappa_{\rm TF}}$, since $C \propto m^*/k_F.$
We point out that there is another condition set by the screening: When $k_F$ is so small that $E_F<\bar\omega$ (entries marked by $\dagger$ in table 1), the screening is no longer static, invalidating  our results which are based on this assumption. The dynamical screening  case has been treated by Riolo et al.~\cite{riolo2025polariton}

We see from the table that for hBN, $g_0$ and $\lambda_0$ are very small ($<$ 0.1) in all cases where the theory is valid. The Bi$_2$Se$_3$ case is more promising in that $g_0$ can reach $\approx0.35$. This value is still not large enough to give significant $T_c$ once the typical $\mu*\approx0.2$ is included, but for metals such as NbSe$_2$ which are already superconducting, the addition contribution can significantly increase $T_c$, especially for the case of Bi$_2$Se$_3$  where $\bar{\omega}$ is comparable to the Debye frequency. For hBN, $\bar{\omega}$ is quite large so that it is between the typical Debye frequency and the Fermi energy. The contribution from the HM may be viewed as a reduction of $\mu^*$ which can lead to a small improvement in $T_c$.

\begin{table}[t]
\centering
\setlength{\tabcolsep}{3pt}
\footnotesize
\begin{tabular}{ccc|cc|cc}
\hline\hline
 & & & \multicolumn{2}{c|}{hBN ($K=1.07$)} & \multicolumn{2}{c}{Bi$_2$Se$_3$ ($K=0.17$)}\\
$k_F$ & $k_Fz$ & $E_F$ & $g_0$ & $\lambda_0$ & $g_0$ & $\lambda_0$\\
(nm$^{-1}$) & & (meV) & & & & \\
\hline
10 & 4.0 & 1905 & 0.034 & 0.026 & 0.065 & 0.061\\
5  & 2.0 & 476  & 0.070 & 0.043 & 0.131 & 0.119\\
3  & 1.2 & 171  & $0.121^{\dagger}$ & $0.059^{\dagger}$ & 0.223 & 0.193\\
2  & $0.8^{*}$ & 76 & $0.190^{\dagger}$ & $0.070^{\dagger}$ & 0.345 & 0.281\\
1  & $0.4^{*}$ & 19 & $0.353^{\dagger}$ & $0.075^{\dagger}$ & $0.685^{\dagger}$ & $0.444^{\dagger}$\\
\hline\hline
\end{tabular}
\begin{flushleft}
\vspace{2pt}
$^{*}$\,$k_Fz<1$: the momentum transfer allowed by the mode profile is no longer confined to a forward window on the Fermi surface. We expect $g_0$ to saturate when $k_Fz\ll 1$. Its expression is given in the text.\\
$^{\dagger}$\,$E_F<\bar\omega$: the static screening used in the text is not justified.
\end{flushleft}
\caption{Dependence on the carrier density at fixed band mass $m^*=2m_e$ and $z=0.4$~nm. We assume a parabolic band so that $v_F=\hbar k_F/m^*$ and $E_F=\hbar^2k_F^2/2m^*$. All entries are calculated for metal sandwiched between HM material on both sides with a gap size equal to $2z$. Here we focus on the contribution of type 2 hyperbolic mode, for which $\bar\omega=185\,$meV in hBN and $13\,$meV in Bi$_2$Se$_3$.  At fixed mass the screening momentum $\kappa_{TF}=(e^2/4\pi\epsilon_\infty)m^*/\pi\hbar^2$ is independent of $k_F$, so that $K=\kappa_{TF}z$ is fixed. Only the parameter $\alpha\propto1/v_F$ and the constant $C\propto1/k_F$ vary in the table. The hBN entries leave their range of validity at $k_F=3.2\,$nm$^{-1}$ while those for Bi$_2$Se$_3$ remain inside down to $k_F=1.3\,$nm$^{-1}$.}
\label{tab:kF}
\end{table}

\section{The self energy of a relativistic boson coupled to HM and application to systems near the Mott insulator.}
\label{boson}
The next problem we address is a charged relativistic boson field $\Psi$ coupled to the HM field in the Coulomb gauge. 
Our motivation is that in the Florence and Georges \cite{florens2004slave} treatment of the  Hubbard model (we consider the 2D case), the charge degree of freedom is treated by a 2D relativistic boson with the dispersion $\pm E_\vec{k}$ where 
\be
E_\vec{k}=\sqrt{v^2\vec{k}^2 +m_b^2}
\ee
so that the upper branch represents the doublon (doubly occupied sites) and the lower branch the holon (empty sites). Here $v$ is the boson velocity and $m_b$ is the mass gap. We calculate the self energy of the boson so the retarded Green function takes the form 
\be \label{boson green function 1}
G^R(\Omega,\vec{k})=\frac{1}{(\Omega+i\delta)^2-E_k^2 -\Sigma^R(\Omega,\vec{k})}
\ee
The Coulomb-gauge Lagrangian is
\begin{equation}
\mathcal L=
\left|\left(\partial_t+i e\phi\right)\Psi\right|^2
-v^2|\bm{\nabla}\Psi|^2-m_b^2|\Psi|^2.
\label{eq:LB}
\end{equation}
The bosons couple to the scalar potential of HM through 
\begin{equation}
\mathcal L^{(1)}
=i e\phi\left[(\partial_t\Psi^*)\Psi-\Psi^*\partial_t\Psi\right],
\qquad
\mathcal L^{(2)}=e^2\phi^2|\Psi|^2.
\end{equation}
If the HM line carries frequency $\nu$, the three-point vertex is 
$
(2\Omega-\nu)g_{\vq\kappa}.
$
The $\phi^2|\Psi|^2$ term contributes at this order through an $\Omega$- and $\vk$-independent tadpole. 
The retarded HM propagator is
\begin{equation}
D^R_{\vq\kappa}(\nu)
=\frac{2\omega_{\vq\kappa}}
{(\nu+i0^+)^2-\omega_{\vq\kappa}^2}
\label{eq:DRB}
\end{equation}
At zero temperature, 
carrying out the internal-frequency integral gives
\begin{align}
\Sigma^R&(\Omega,\vk)
=\int \frac{d^2 q}{(2\pi)^2}\int_0^\infty\frac{d\kappa}{2\pi}|g_{\vq\kappa}|^2
\\
&\bigg[
\frac{(2\Omega-\omega_{\vq\kappa})^2}
{2E_{\vk-\vq}\left(\Omega-E_{\vk-\vq}-\omega_{\vq\kappa}+i0^+\right)}
\nonumber\\[-1mm]
\hspace{11mm}
&-\frac{(2\Omega+\omega_{\vq\kappa})^2}
{2E_{\vk-\vq}\left(\Omega+E_{\vk-\vq}+\omega_{\vq\kappa}+i0^+\right)}
+\frac{\omega_{\vq\kappa}}{E_{\vk-\vq}}
\bigg].
\label{eq:SigmaB}
\end{align}
In the low-energy region used below, neither denominator reaches its decay threshold, so $\operatorname{Im}\Sigma^R=0$ and Eq.~\eqref{eq:SigmaB} directly gives $\operatorname{Re}\Sigma^R$.

Let $\Sigma_0$ denote the frequency and  momentum independent mass shift, including the tadpole and the constant part of the exchange diagram.  The low-energy expansion is
\begin{equation}
{
\operatorname{Re}\Sigma^R(\Omega,\vk)
=\Sigma_0+B_\Omega\Omega^2+B_kv^2k^2+\cdots.
}
\label{eq:SigmaB-expansion}
\end{equation}
At $\Omega=0$,
\begin{equation}
\operatorname{Re}\Sigma^R(0,\vk)
=\int \frac{d^2q}{(2\pi)^2}
\int_0^\infty\frac{d\kappa}{2\pi}|g_{\vq\kappa}|^2
\frac{\omega_{\vq\kappa}}{E_{\vk-\vq}+\omega_{\vq\kappa}}.
\label{eq:SigmaBzero}
\end{equation}
Expanding Eqs.~\eqref{eq:SigmaB} and \eqref{eq:SigmaBzero} gives the  coefficients
\begin{equation}
B_\Omega=
-\int\!\frac{d^2q}{(2\pi)^2}\int_0^\infty\frac{d\kappa}{2\pi}|g_{\vq\kappa}|^2
\frac{(2E_{\vq}+\omega_{\vq\kappa})^2}{E_{\vq}(E_{\vq}+\omega_{\vq\kappa})^3}
\label{eq:COmega}
\end{equation}
and
\begin{equation}
B_k=
\int\!\frac{d^2q}{(2\pi)^2}\int_0^\infty\frac{d\kappa}{2\pi}|g_{\vq\kappa}|^2
\frac{\omega_{\vq\kappa}
\left[E_{\vq}^3-\omega_{\vq\kappa}E_{\vq}^2
-3m_b^2E_{\vq}-m_b^2\omega_{\vq\kappa}\right]}{4E_{\vq}^3(E_{\vq}+\omega_{\vq\kappa})^3}.
\label{eq:CkB}
\end{equation}
{For poles satisfying $m_b',v'k\ll\bar\omega$, the expansion in Eq.~\eqref{eq:SigmaB-expansion} gives}
\begin{equation}
G^R(\Omega,\vk)
\simeq
\frac{Z}{(\Omega+i0^+)^2-v'^2k^2-m_b'^2},
\end{equation}
with
\begin{equation}
Z=\frac{1}{1-B_\Omega},
\qquad
\frac{v'^2}{v^2}=\frac{1+B_k}{1-B_\Omega},
\qquad
m_b'^2=\frac{m_b^2+\Sigma_0}{1-B_\Omega}.
\label{eq:ZvB}
\end{equation}
The critical point is fixed by $m_b^2+\Sigma_0=0$; in the infrared formulas below, $m_b'^2$ denotes the corresponding distance from criticality.

The results presented below is for the case where the system sits on top of the HM material. For this one-sided geometry,
$
|f(\vq,\kappa,z)|^2
=f_{\vq\kappa}^{\,2}e^{-2qz}
$, and
\begin{equation}
\int_0^\infty\frac{d\kappa}{2\pi}|g_{\vq\kappa}|^2
=\frac{e^2\bar\omega}{4\sqrt2\epsilon_\infty q}
\frac{\eta^2}{\omega_T\omega_L}
\sqrt{\frac{\omega_T}{\omega_L}}
 e^{-2qz}.
\label{eq:g2B}
\end{equation}
We define
$
\alpha_b=\frac{v}{z\bar\omega}.
$
At the critical point, $m_b'=0$, (we will drop the $'$ label from this point on) and the integrals in Eqs.~\eqref{eq:COmega} and \eqref{eq:CkB} are logarithmically divergent. To leading order in $m_b/\bar{\omega}$, we introduce an infra-red cut-off $x_{\rm IR} = \max\!\left(\frac{m_b}{\bar\omega},\frac{v}{\widetilde c}
\right)$. Since $\frac{v}{\widetilde c}$ is very small, we can set $x_{\rm IR}$ to be   $m_b/\bar{\omega}$ in most cases of interest. We obtain 
\begin{equation}
B_\Omega=-
\frac{e^2}{8\sqrt2\pi\epsilon_\infty v}
\frac{\eta^2}{\omega_T\omega_L}
\sqrt{\frac{\omega_T}{\omega_L}}
\int_{x_{\rm IR}}^{\infty}dx\,
 e^{-2x/\alpha_b}
\frac{(2x+1)^2}{x(x+1)^3}
\label{eq:COmegaReduced}
\end{equation}
\begin{equation}
B_k=
\frac{e^2}{32\sqrt2\pi\epsilon_\infty v}
\frac{\eta^2}{\omega_T\omega_L}
\sqrt{\frac{\omega_T}{\omega_L}}
\int_{x_{\rm IR}}^{\infty}dx\,
 e^{-2x/\alpha_b}
\frac{x-1}{x(x+1)^3}
\label{eq:CkReduced}
\end{equation}
Here $x=vq/\bar\omega$. 
Consequently for $\alpha_b \ll 1$,
\begin{equation}
B_\Omega=-
C_v \begin{cases}
\log \frac{{\rm min}(1,\alpha_b)}{x_{\rm IR}}, &{x_{\rm IR}}\ll \alpha_b\ll 1\\
\frac{\alpha_b}{2x_{\rm IR}}e^{-2x_{\rm IR}/\alpha_b} , & \alpha_b \ll x_{\rm IR}\ll 1
\end{cases}
\label{eq:COmegaReduced2}
\end{equation}
\begin{equation}
B_k=B_\Omega/4, \quad x_{\rm IR} \ll 1
\label{eq:CkReduced2}
\end{equation}
where $C_v$ is defined in Eq.~\ref{C_v}.    The conditions in the two lines in Eq.~\ref{eq:COmegaReduced2} becomes $v/z > {\rm or} < m_b$ respectively.  For large $z$ so that $\frac{v}{z\bar{\omega}}<1 $, the second line 
applies and $B_\Omega$  decays exponentially with $z$. 
On the other hand, for very small $z$ so that $\alpha_b \gg 1$, the integral for  $B_\Omega$ is of order unity while $B_k$ may change sign for large enough $x_{\rm IR}$.
 
To summarize, the $Z$ factor and the velocity renormalization are given by

\label{eq:vBfinal1}
\be
Z\simeq (1-B_\Omega)^{-1}
\ee
\be
\frac{v'^2(k)-v^2}{v^2}
\simeq \frac{5}{4}B_\Omega, \quad \alpha_b<1,   \, x_{\rm IR} < 1.
\label{eq:vBfinal2}
\ee
We can repeat this calculation for the type 1 HM. The only place where this differs from type 2 HM is in the normalization factor which we now call $f_{1,\vq\kappa}^2$. It is given in Eq.\ref{f kappa 1} and behaves quite differently from $f_{\vq\kappa}^2$. 
It turns out that the results for the  $\Omega^2$ and $k^2$ expansion of  the self energy remains the same, as long as the parameters in Eqs.\ref{eq:COmegaReduced} and \ref{eq:CkReduced} are replaced by the corresponding entities for the type 1 HM. 

We  now briefly review the formulation of Florence and Georges \cite{florens2004slave} which describe the metal-insulator transition for the Hubbard model at half-filling. The electron operator is written as a product of a rotor and a fermion $f_{\sigma}$ that carries spin and no charge. The rotor is in turn approximated by a relativistic boson field $X$ that carries the charge degree of freedom. The boson must satisfy a constraint $X^{\dagger}X=1$ on each site which is enforced on average by the following equation.
\be \label{boson eq}
1=T\sum_{\omega_n}\int \frac{d\vec{q}}{(2\pi)^2}\frac{1}{\omega_n^2/U + \lambda +Q_X \epsilon(q)}
\ee
The term inside the integral is the $X$ boson Green function in Matsubara space, where $\lambda$ is a Lagrangian multiplier to be solved to satisfy this constraint equation, $\epsilon(q)$ is a tight binding band dispersion that is proportional to the  hopping $t$. For example, for nearest neighbor hopping on a square lattice with lattice constant $a_b$, $\epsilon(q)=-2t(\cos(q_x a_b)+\cos(q_y a_b)$). $Q_X$ is a parameter determined  from solving a mean field equation. 
As shown by Florens and Georges \cite{florens2004slave}  at zero temperature,  $Q_X$ is simply given by a numerical constant. From the pole of the Green function, the boson dispersion is given by  $E_X(q)=\pm \sqrt{UQ_X\epsilon(q)+U\lambda}$. It is convenient to introduce  $D_0=4t$ which is half of the $\epsilon(q)$ bandwidth and shift the energy by $D_0$ so that the band bottom is at zero energy.  We now write $E_X(q)=\pm \sqrt{UQ_X(\epsilon(q)+D_0)+m_X^2}$ where $m_X^2=U(\lambda-Q_X )D_0$. This dispersion is characterized by a Dirac-like spectrum with an energy gap $2m_X$ and describe a Mott insulator with a charge gap. The upper and lower branches are the doublon and holon bands. Note that $m_X^2$ now takes over the role of the Lagrange multiplier $\lambda$ and its value is determined by solving the constraint equation Eq.\eqref{boson eq}. We perform the Matsubara sum in Eq.\eqref{boson eq}  by converting it to an integral at $T=0$ and  obtain
\be 
\label{boson eq 1}
1=\int \frac{d\vec{q}}{(2\pi)^2}\frac{U}{E_X(q)}=\int \frac{d\vec{q}}{(2\pi)^2}\frac{U}{\sqrt{UQ_X(\epsilon(q)+D_0)+m_X^2}}
\ee
As $U$ is reduced, the RHS of this equation decreases and at some critical value $U=U_c$,   Eq.\ref{boson eq 1} can no longer be satisfied without Bose condensation to the $q=0$ state. This point is set by $m_X=0$ and  marks the transition from the Mott insulator to a metal. 
Once the boson condenses,  the fermion acquires the charge and turns into an electron band to form a metal. Notably, the uncondensed part of the  boson spectrum remains and the conduction band emerges at zero energy, initially with a small spectral weight. This result predicts that for the metal near the Mott transition, the Drude weight is small and is accompanied by a broad infra-red absorption peak which can be thought of as transition between the central peak and  remnants of the upper and lower Hubbard bands as well as between the Hubbard bands themselves. Optical experiments have found that in $\kappa-ET$ salts, a broad mid-infra red peak must be added to the Drude peak to fit the data, consistent with this picture.\cite{dressel2004optical}

The critical $U_c$ is solved by setting $m_X=0$ in Eq.\ref{boson eq 1}. Note that the RHS is proportional to $\sqrt{U/t}$ and the solution for $U_c $ can readily be found.

We will now describe how the the critical $U_c$ is affected by coupling to the HM mode of hBN by  apply our results on the relativistic boson to the $X$ boson described by Eq.\eqref{boson eq}. We can expand the first term inside the square root in Eq.\eqref{boson eq 1} which goes as $q^2$ at the bottom of the band and replace it by $v^2q^2$. The $X$ field Green function now looks the same as that for the $\Psi$ boson except for a factor $U$ in the numerator which is a matter of convention. As discussed earlier, the coupling to the HM mode adds a $Z $ factor in the numerator of the Green function and convert the velocity from $v$ to $v'$. Setting $m_X=0 $ in Eq.\eqref{boson eq 1}, we see that the RHS now has an extra factor of $Z/(v'/v)$. Since the RHS is proportional to $\sqrt{U}$, we find a new critical $\tilde{U_c}$ which is given by $\sqrt{\tilde{U_c}}Z/(v'/v)=\sqrt{U_c}$. Using Eqs.\ref{eq:ZvB} we conclude that 
\be \label{Uc}
\tilde{U_c}=U_c(1+B_k)/Z
\ee
The negative $B_k$ term decreases the critical $U$ while the $Z$ factor, being less than unity moves it in the opposite direction. Since $B_k\approx B_\Omega/4$ in many cases, the $Z$ factor dominates , so that the critical $U$ is increased due to the coupling to the HM mode: the system has been driven in a more metallic direction. A simpler way to see this is that if the RHS of Eq.~\eqref{boson eq 1} is multiplied by $Z < 1$, a smaller $m_X$ (i.e. smaller charge gap)  is required to satisfy the constraint equation. While our calculation was carried out on the insulating side, the same trend is expected on the metallic side.  For the organic system, this tendency is indicated by the red arrow in the phase diagram shown in Fig.\ref{fig:phase diagram.jpg}. 
While in principle a superconductor may be pushed outside the superconducting regime, the experiment of Keren et al.~\cite{keren2026cavity} requires an effect over hundreds of nanometers and cannot be explained by our effect which decays 
rapidly with $z$ for large $z$.
On the other hand, it will be interesting to start with an insulator close to the Mott transition and see if the top layer can be made superconducting. Possible applications to experiments will be described in the discussion section. 


\section{Gauge dependence of Z factor}

The calculation of the previous section is carried out in the Coulomb gauge which incorporate the longitudinal field of the HM in the scalar potential $\phi$. One may be tempted to use another gauge, such as the temporal (or Weyl) gauge, which sets $\phi=0$, but keep both the longitudinal and transverse parts of $\vec{A}$. For calculations of physical quantities, such as the velocity renormalization the results must agree. However they do not agree for quantities such as the Z factor. This is because the standard electron Green function $G(\vk,t)=-i\big\langle T\, c_{\vk}(t)\,c^\dagger_{\vk}(0)\big\rangle$ which contains only the electron operators  is not gauge invariant. What we need to do is to compute physical quantities such as the spectral weight of a quasi-particle measured in ARPES experiment or the reduction in the tunneling DOS in a tunneling experiment. We shall argue that the Z factor computed in the Coulomb gauge is the one that gives the correct result for the physical quantities being measured. In earlier unpublished works, 
we had worked in the temporal (Weyl) gauge  and evaluated the self energy with the standard electron operator. That calculation (reproduced in Appendix~\ref{appendix: weyl gauge}) gives a parametrically larger $B_\Omega=1-Z$, with a different dependence on the distance to the slab. 
These results are incorrect. On the other hand, the velocity correction is gauge invariant and indeed there is a cancellation between $B_\Omega$ and $B_k$ so that the parametrically larger term does not appear in the velocity correction. However, this computation is more combersome and the Coulomb gauge is the preferred method. In this section we explain the issue of gauge choice in more details.

Since the electron is always interacting with the gauge field $\mathbf{A}$, it propagation from $\mathbf{x}$ to $\mathbf{x'}$ is accompanied by  To the unitary operator
\begin{equation}
U_{\rm A}=\exp\left[ie\int_{\mathbf x}^{\mathbf x'}\mathrm{d}\boldsymbol\ell\cdot\mathbf A\right].
\end{equation}
  A physical way to understand this is by the Feynman path integral formulation, which  states that the propagation of an electron is modified by the line integral $e\int_{\mathbf x}^{\mathbf x'}\mathrm{d}\boldsymbol\ell\cdot\mathbf A$.

Next we decompose  $\mathbf{A}$ into its longitudinal and transverse components:
$\mathbf A=\mathbf A_{\parallel}+\mathbf A_{\perp}$. The longitudinal  part is always the gradient of some operator, $\mathbf A_{\parallel}=\boldsymbol\nabla\chi$; the transverse part is not. The operator $U_{\rm A}$ splits accordingly:
\begin{equation}
\exp\left[ie\int_{\mathbf x}^{\mathbf x'}\mathrm{d}\boldsymbol\ell\cdot\mathbf A\right]
=e^{ie[\chi(\mathbf x')-\chi(\mathbf x)]}
\times
\underbrace{\exp\left[ie\int_{\mathbf x}^{\mathbf x'}\mathrm{d}\boldsymbol\ell\cdot\mathbf A_{\perp}\right]}_{\text{path-dependent}},
\end{equation}
The transverse part is gauge invariant and path dependent, and its effect is captured by terms like $\mathbf{A_{\perp}} \cdot \bf{j}$ in the coupling Hamiltonian in the usual way. We therefore define a gauge invariant electron Green function by inserting 
\begin{equation}
U_{\rm A_{\parallel}}=\exp\left[ie\int_{\mathbf x}^{\mathbf x'}\mathrm{d}\boldsymbol\ell\cdot\mathbf A_{\parallel}\right].
\end{equation}
between the electron creation and annihilation operators in the standard real space electron Green function.
The longitudinal part is gauge dependent and its contribution needs to be kept for a general gauge. The Coulomb gauge is special in that this term is absent. Instead the potential is given by the solution of the Poisson equation for a given charge configuration. Therefore the electron Green function defined in the standard way coincides with the gauge invariant Green function defined above in the Coulomb gauge. The $Z$ factor calculate in the Coulomb gauge corresponds to the spectral weight of a physical excitation. 
We think it is instructive to carry out a calculation of the gauge invariant Green function in the Weyl gauge and show that the resulting quasi-particle spectral weight and Z factor coincides with a calculation of the stand Green function using the Coulomb gauge. This is presented  in Appendix \ref{appendix: weyl gauge}.

\section{Discussion}
We find that HM can produce a enormous zero point fluctuation in the electric field which can have an  effect on the electronic properties of a proximate material, provide the distance $z $ is close enough, of order a few  atomic layers. In the case of coupling to a Fermi liquid, the coupling strength is set by the parameter $C$ which us a product of the dimensionless number $\frac{e^2}{4\pi\epsilon_\infty v_F}$ which depends on the metal and a number which depends on the  polarizability of the HM  material. It further depends on two dimensionless couplings: $\alpha=v_F/z\bar{\omega}$ and $K=\kappa_{TF}z$, which depends on the distance between the metal and the HM material. We find that the effect is quite small for distance $z$ that is achievable. 
Generally speaking, the effect is largest for small Fermi velocity and low density. In these cases, we predict a suppression of the Z factor which may be observable in STS experiment as a dip in the tunneling DOS below the average HM scale of $\bar{\omega}$. Contribution to pairing is typically too weak to overcome the effective repulsion $\mu^*$, but for an existing superconductor such as NbSe$_2$, proximity to HM may enhance the transition temperature. 

We find that the effective interaction due to the exchange  of the HM is not that different from the exchange of an isotropic polar phonon treated in the Fr\"ohlich Hamiltonian. Indeed, the enhancement of $T_c$ when the iron based superconductor FeSe is placed next to SrTiO$_3$ is perhaps the only example where such an effect has been documented.~\cite{lee2014interfacial} The difficulty is that it is difficult to place a dielectric very close to a metal. From this point of view, the van der Waal system of metals and dielectric is particularly attractive, because it is possible to place these materials very close to each other without direct modifications of   physical properties such as charge transfer. Since these dielectrics are intrinsically anisotropic, the presence of HM is the rule rather than the exception, with hBN and Bi$_2$Se$_3$ being particularly good example. Therefore the results of our study are relevant to realistic experimental set-ups and hopefully will  lead to further experimental investigations.

We next focus on systems near the Mott transition and we find that HM can modify the strength of the correlation                                                                                                                                                                                                                                                                                                                                                                                                                                                                                                                                                                                                                                                                                                                                                                                                                                                                                                                                                                                                                                                                                                                                                                                                                                                                                                                                                                                                                                                                                                                                                                                                                                                                                                                                                                                                                                                                                                                                                                                                                                                                                                                                                                                                                                                                                                                                                                                                                                                                                                                                                                                                                                                                                                                                                                                                                                                                                                                                                                                                                                                                                                                                                                                                                                                                                                                                                                                                                                                                                                                                                                                                                                                                                                                                                                                                                                                                                                                                                                                                                                                                                                                                                                                                                                                                                                                                                                                                                                                                                                                                                                                                                                                                                                                                                                                                                                                                                                                                                                                                                                                                                                                                                                                                                                                                                                                                                                                                                                                                                                                                                                                                                                                                                                    and push the system to be more metallic. While our investigation is inspired by the work Keren et al.\cite{keren2026cavity} their analysis indicates that the effect of the hBN must extends hundreds of nanometer into the organic superconductor in order to produce the signal they observe. The effect we find decreases rapidly with distance, and we cannot offer an explanation of their observation. Nevertheless, we would like to consider situations where an observable effect is possible and provide some estimates for realistic materials. 

In view of the recent experiment\cite{keren2026cavity}, we begin with considering the $\kappa-ET$ system of organic compounds and focus on the first monolayer, i.e., we set $z$ to be a typical layer spacing. Based on the phase diagram in Fig.\ref{fig:phase diagram.jpg}, the idea is to start on the insulating side and see if the hBN can push the first few layers to become metallic or superconducting. Good candidates are the AF compound $\kappa-ET-(CN)_2Cl$ and the spin liquid candidate $\kappa-ET-(CN)_3$ which are known to be very close to the Mott transition.~\cite{dressel2004optical} 
Meissner effect or transport experiment may be able to detect the conversion of the top few layers from  insulator to metal or superconductor, as opposed to the more difficult task of detecting the suppression of the Meissner signal of a few layers.  
Another interesting candidate is the  SC $\kappa-$(ET)I$_3$ which has a relatively low $T_c$ and one can see whether $T_c $ will increase or decrease. 

With this motivation we now make some rough estimate of the size of the effect we may expect for the first layer. We would like to estimate the percentage change of the critical $U$ that controls the metal insulator transition. Our theory indicates that it is given by Eq.\ref{Uc} and we will now estimate the size of $B_\Omega$. 
We begin with an estimate of the boson velocity $v$. Taking the tight binding model for a square lattice for simplicity, we write $v=t_ba_b$ where $t_b$ is the effective boson hopping and $a_b $ is the lattice constant which we take to be the distance between ET dimers which is  $\approx 0.8 nm$. We write the half bandwidth $D=4t_b$ resulting in $v=D a_b/4$. While band structure calculations find a bare hopping $t_0$ to be about 50 meV, the bandwidth of the X boson band has a suppression factor $Q_X$ which is difficult to calculate accurately. Instead we extract $D$ from the optical absorption experiment. As mentioned earlier the experimentalists observe an IR peak with a width of about 2000 $cm^{-1}$ that extends beyond the Drude peak which is interpreted as originating from transition between a Drude peak and the boson band.\cite{dressel2004optical} We expect the width of this peak to be roughly the full boson bandwidth which is 2D. Therefore we set $D\approx 1000\,cm^{-1}$. This is a factor $1.6$ 
smaller than $4t_0$ and is reasonable. With this value 
we find $c/v \approx 8\times 10^{3} $ which is large, but not unreasonable for this narrow band system.

\begin{table*}[htbp]
\begin{ruledtabular}
\begin{tabular}{llcccccccc}
system & substrate & mode & $\bar\omega$ & $C_v$ & $\alpha_b$ & $x_{\rm IR}$ & $B_\Omega$ & $B_k$ & $\tilde U_c/U_c$\\
 & & & $\frac{\omega_T+\omega_L}{2}$
 & Eq.~\ref{C_v}
 & $\frac{v}{z\bar\omega}$
 & $\frac{m_b}{\bar\omega}$
 & Eq.~\ref{eq:COmegaReduced}
 & Eq.~\ref{eq:CkReduced}
 & $\frac{1+B_k}{Z}$\\
 & & & (meV) & & & & & & $Z=\frac{1}{1-B_\Omega}$\\
\hline
$\kappa-(CN)_2Cl$   & hBN                & type 2 & 185 & 1.33 & 0.17 & 0.17 & $-0.074$ & $-0.0069$ & \\
$v=24.8$\,meV\,nm   &                    & type 1 & 100 & 0.57 & 0.31 & 0.31 & $-0.033$ & $-0.0015$ & \\
$m_b=31$\,meV       &                    & sum    &     &      &      &      & $-0.107$ & $-0.0083$ & $1.10$\\
\cline{2-10}
$z=0.8$\,nm         & Bi$_2$Se$_3$       & type 2 & 13  & 0.86 & 2.39 & 2.39 & $-0.031$ & $+0.0003$ & \\
                    &                    & type 1 & 19  & 0.11 & 1.65 & 1.64 & $-0.005$ & $+0.0001$ & \\
                    &                    & sum    &     &      &      &      & $-0.036$ & $+0.0004$ & $1.04$\\
\hline
1T-TaSe$_2$         & hBN                & type 2 & 185 & 1.06 & 0.42 & 0.29 & $-0.144$ & $-0.0063$ & \\
$v=31$\,meV\,nm     &                    & type 1 & 100 & 0.45 & 0.78 & 0.55 & $-0.060$ & $-0.0006$ & \\
$m_b=54.5$\,meV     &                    & sum    &     &      &      &      & $-0.204$ & $-0.0070$ & $1.20$\\
\cline{2-10}
$z=0.4$\,nm         & Bi$_2$Se$_3$       & type 2 & 13  & 0.69 & 5.96 & 4.19 & $-0.039$ & $+0.0003$ & \\
                    &                    & type 1 & 19  & 0.09 & 4.11 & 2.89 & $-0.006$ & $+0.0001$ & \\
                    &                    & sum    &     &      &      &      & $-0.046$ & $+0.0004$ & $1.05$\\
\hline
1T-TaS$_2$          & hBN                & type 2 & 185 & 1.06 & 0.42 & 0.54 & $-0.028$ & $-0.0004$ & \\
$v=31$\,meV\,nm     &                    & type 1 & 100 & 0.45 & 0.78 & 1.00 & $-0.011$ & $+0.0001$ & \\
$m_b=100$\,meV      &                    & sum    &     &      &      &      & $-0.039$ & $-0.0004$ & $1.04$\\
\cline{2-10}
$z=0.4$\,nm         & Bi$_2$Se$_3$       & type 2 & 13  & 0.69 & 5.96 & 7.70 & $-0.005$ & $+0.0000$ & \\
                    &                    & type 1 & 19  & 0.09 & 4.11 & 5.31 & $-0.001$ & $+0.0000$ & \\
                    &                    & sum    &     &      &      &      & $-0.006$ & $+0.0000$ & $1.01$\\
\end{tabular}
\end{ruledtabular}
\caption{Estimates for the three candidate systems on the two substrates, one hyperbolic mode per line.  For each system the boson velocity $v=Da_b/4$, the gap $m_b$ and the distance $z$ are those quoted in the text.  Each column is defined in its header; $B_\Omega$ and $B_k$ are obtained by evaluating Eqs.~\ref{eq:COmegaReduced} and \ref{eq:CkReduced} numerically rather than from their asymptotic forms, which is why $B_k$ is positive whenever $x_{\rm IR}>1$ and why the ratio $B_k/B_\Omega$ reaches $1/4$ only when $x_{\rm IR}\ll1$.  The two modes' contributions add, and the last column gives $\tilde U_c/U_c=(1+B_k)/Z$ with $Z=1/(1-B_\Omega)$.  
Here we have used the following values for $\omega_T$ and $\omega_L$ in each system: for hBN, $1370\,cm^{-1}$ and $1611\,cm^{-1}$ for the type 2 HM, and $777\,cm^{-1}$ and $829\,cm^{-1}$ for the type 1 HM; for Bi$_2$Se$_3$, $64\,cm^{-1}$ and $146\,cm^{-1}$ for the type 2 HM, and $146\,cm^{-1}$ and $158\,cm^{-1}$ for the type 1 HM.  The type 2 window is opened by the in plane phonon and the type 1 window by the out of plane one.
The entries for $B_\Omega$ and $B_k$ are for the one-sided geometry and double if the layer is encapsulated. 
}
\label{tab:estimates}
\end{table*}

Putting this velocity into the expression for $C_v$ and using the parameters for hBN given in the introduction, we find $C_v \approx 1.33$.

Next we estimate the parameter $\alpha_b=v/z\bar{\omega}$. 
If we take $z=0.8 nm$ we find $\alpha=0.18$ 
For $\kappa-(CN)_2Cl$ the Mott gap $m_b$ (half the optical gap) is estimated to be $250 cm^{-1}$.~\cite{elsasser2012power}. We find $m_b/\bar{\omega} \approx 0.18$ as well
. This places us in the cross-over regime between the two lines in Eq.~\ref{eq:COmegaReduced2}.  The exponential cuts the integral at $x\simeq\alpha_b/2=0.09$, which is slightly below $x_{\rm IR}=0.18$.
We evaluate $B_k$ directly from Eq.~\ref{eq:COmegaReduced2} which gives $B_\Omega=-0.075$.
Apparently $x_{\rm IR}=0.18$ is not small enough for  
Eq.~\ref{eq:CkReduced2} to apply: 
 direct evaluation gives $B_k=-0.006$, that is $B_k/B_\Omega=0.09$ rather than $1/4$.  With $Z=1/(1-B_\Omega)=0.93$, Eq.~\ref{Uc} gives $\tilde U_c/U_c=(1+B_k)/Z=1.07$.  The $z$ dependence is the exponential $e^{-2m_bz/v}$ of the second line, of decay length $v/2m_b=0.4\,$nm, so the effect is confined to the first layer.  

We repeat these steps for the type 1 HM and find that its contribution is comparable. 

We consider another example  Bi$_2$Se$_3$ where a strongly coupled type 2 HM mode exists with the following parameters: $\omega_L=146, cm^{-1}$, $\epsilon_\infty=29$, $\eta^2/\omega_L \omega_T=1.84$ and $\sqrt{\omega_T/\omega_L}=0.66$. \cite{wu2015topological}. We find $C_v \approx 0.86$.
The mode frequency is about 10 times smaller than in hBN.
With $\omega_L=146\,cm^{-1}$ and $\sqrt{\omega_T/\omega_L}=0.66$ we have $\omega_T=64\,cm^{-1}$ and $\bar\omega=105\,cm^{-1}=13\,$meV.  At the same $z=0.8\,$nm this makes $\alpha_b=v/z\bar\omega=0.0248/(0.8\times0.013)=2.4$, so that $\alpha_b/2=1.2$. But the same small $\bar\omega$ raises the infrared cutoff to $x_{\rm IR}=m_b/\bar\omega=31/13=2.4$, which is larger than unity: the Mott gap of the organic now exceeds the HM frequency and the logarithmic window is closed altogether.  Evaluating Eq.~\ref{eq:COmegaReduced} directly gives $B_\Omega=-0.031$ and $B_k=+0.0003$, so that $Z=0.97$ and $\tilde U_c/U_c=1.03$. 
Another   interesting class of materials in the 1T-TaS$_2$, 1T-TaSe$_2$ and 1T-NbSe$_2$ family.  These materials are  close to a Mott transition and may exhibit spin liquid states \cite{law20171t}. An important advantage is that these materials are available in monolayers and can be probed by a variety of techniques including scanning tunneling spectroscopy (STS) and angle resolved photoemission (ARPES). The Mott insulator is associated with clusters of 13 atoms that form a star of David charge density wave pattern. Hence the unit cell size is rather large, $a_b=1.24\, nm$,  giving  a relatively small velocity. STS can be done on a monolayer deposited on hBN and  direct information on the energy spectrum  can be obtained. For example, 1T-TaSe$_2$ on oriented graphite shows an energy gap $2\Delta \approx 109\, meV$ and we can estimate $2D\approx200\, meV$ from the width of the upper or lower Hubbard bands.\cite{ruan2021evidence} Another interesting case is 1T-NbSe$_2$ which is metallic in the bulk but the monolayer was found to be a Mott insulator.\cite{nakata2016monolayer} On the other hand, 1T-TaS$_2$ shows a larger gap of $2\Delta \approx 200 meV$.\cite{vavno2021artificial} Interestingly bulk 1T-TaS$_2$ becomes a 5K SC under pressure \cite{sipos2008mott, ritschel2013pressure}, reminiscent of the appearance of SC near at the Mott insulator boundary seen in the organic compounds and more recently in twisted WSe$_2$.\cite{xia2026bandwidth,guo2026angle} By placing a monolayer of these materials  on hBN, STS experiment can probe the evolution of the gap and check whether is is increased, decreased or even closed. Using the parameters mentioned above for 1T-TaSe$_2$ and similar to what was shown below Eq.\ref{eq:Tc}, we find $ v=0.031\, {\rm eV nm}$ and  $c/v \approx 6.4 \times 10^3$, $C_{v,2}     
=1.06 $ 
. Setting $z=0.4 nm$ we find 
$\alpha_b=0.031/(0.4\times0.17)=0.46$, and $m_b=\Delta=54.5\,$meV gives $m_b/\bar\omega_{2}=0.32$ for the type 2 HM . 
 $\alpha_{b,2}/2=0.23<x_{\rm IR,2}=0.32$ and the second line applies: the direct evaluation gives $B_\Omega=-0.14$ and $B_k=-0.006$, hence $Z=0.87$ and $\tilde U_c/U_c=1.14$ . 

Next we account for the type 1 HM of hBN.  Its parameters are $\omega_{1T}=777\,cm^{-1}$ and $\eta_1/\omega_{1T}=0.37$, so that $\omega_{1L}=\sqrt{\omega_{1T}^2+\eta_1^2}=829\,cm^{-1}$ and $\bar\omega_1=803\,cm^{-1}=99.5\,$meV.  The numerical factor is $(2\sqrt2)^{-1}(\eta_1^2/\omega_{1T}\omega_{1L})\sqrt{\omega_{1T}/\omega_{1L}}=0.044$, 
so that $C_{v,1}=
0.45$.  The lower frequency works the other way: at the same $z=0.4\,$nm, $\alpha_{b,1}=0.031/(0.4\times0.0995)
=0.78$ and $x_{{\rm IR},1}=54.5/99.5=0.55$, both about twice the type 2 values.  Since $\alpha_{b,1}/2=0.39<x_{{\rm IR},1}=0.55$ the second line again applies, and the direct evaluation of Eq.~\ref{eq:COmegaReduced} gives similar value as for type 2, $0.132$ versus $0.135$. 
Hence $B_{\Omega,1}=-0.060$ and $B_{k,1}=-0.0006$.  Adding the two modes,
\begin{align}
B_\Omega&=-0.144-0.060=-0.204,\nonumber\\
B_k&=-0.0063-0.0006=-0.0070,
\end{align}
so that $Z=1/(1-B_\Omega)=0.83$ and Eq.~\ref{Uc} gives $\tilde U_c/U_c=(1+B_k)/Z=1.20$.  If the monolayer is encapsulated, with hBN on both sides, both contributions double, giving $B_\Omega=-0.41$, $Z=0.71$ and $\tilde U_c/U_c=1.39$. 

In summary, placing systems that are just on the insulating side of the Mott transition in proximity with HM materials may have observable effects on shifting the system towards the metallic side. The organic compounds are interesting candidates to explore.   The 1T-TaSe$_2$ and related family are predicted to show stronger effect and  are particularly promising examples.

\section*{Acknowledgment} 
We thank Itai Keren for the introducion to the subject and to Dmitri Basov, Taige Wang, Andrey Chubukov, Leonid Levitov, Iliya Essin, Marios Michael, Eugene Demler and Angel Rubio for discussions. PL  acknowledges support by DOE (USA) office of Basic Sciences Grant No. DE-FG02-03ER46076. Z. D. acknowledges support from the Gordon and Betty Moore Foundation’s EPiQS Initiative, Grant GBMF8682.

\appendix
\begin{widetext}
\section{The local HM phonon density of states.}
\label{appendix:DOS}

We calculate the photon density of a type 2 HM by expanding near $\omega_L$ and $\omega_T$.
First we consider $\omega_L-\omega$ small and introduce the small difference
\be\label{Delta omega 1}
\Delta_{\omega}^2 = \omega_L^2-\omega^2.
\ee
We assume the condition  $|\tilde{\epsilon}|<<1$, in which case $\tilde{\epsilon}$ is expanded as
\be\label{eps expansion 1}
\tilde{\epsilon}=-\Delta_{\omega}^2/\eta^2
\ee
where $\eta^2=\omega_L^2-\omega_T^2$ and $\tilde{\epsilon}$  vanishes linearly for $\omega$ just less than $\omega_L$ .

We write the photon density of states at a position $z$ outside the hBN slab as
\be\label{DOS def 1}
\rho(\omega, z)= \int_{0}^{\Lambda_q} \frac{d\kappa}{2\pi}\int_{0}^{\Lambda_q} \frac{d\vec{q}}{(2\pi)^2}\,\delta(\omega-\omega_{\vec{q} 
\kappa} ) |f(\vec{q,\kappa,z})|^2
\ee
 We assume the thickness $d>>z$ and take the continuum limit for the $\kappa$ integration to replace the sum over mode indices. The upper limit $\Lambda_q$ is the inverse of a  hBN lattice scale below which the dielectric function approximation is valid.   The local wavefunction is given by Eq.\ref{wf main}
\be \label{wf}
|f(\vec{q,\kappa,z}|^2=f_\vec{q,\kappa}^2(\cos(\kappa z))^2 e^{-2 \nu(q)z}
\ee 
where $\nu_{\vec{q}} \approx q$ is given in Eq.\ref{nu def}. For $\omega_{\vec{q} 
\kappa}$ near $\omega_L$, $\kappa << q$ according to Eq.\ref{q vs kappa} and since $q$ is cut-off by $1/2z$,  we can set $cos(\kappa z)) =1$. 
It is shown in Appendix 2 Eq.\ref{f near L} that for  $\omega $ near $\omega_L$ , $f_\vec{q,\kappa}^2/2$  is well approximated by a constant equal to $\eta^2/\omega_L^2$. Then Eq. \eqref{DOS def} becomes
\be\label{DOS int1 appendix}
\rho(\omega)= \frac{\eta^2}{\omega_L^2}\int_{0}^{\Lambda_q} \frac{d\kappa}{\pi}\int_{\kappa}^{\Lambda_q} \frac{d(q^2)}{2\pi}\,\delta(\Delta_{\omega}^2-\Delta_\vec{q}^2) \omega  e^{-2qz}
\ee
where we define 
\be\label{delta q 1}
\Delta_\vec{q}^2=\omega_L^2-\omega_{\vec{q}\kappa}^2 
\ee
which is similar to Eq. \ref{Delta omega} but is  a function of $q$ for fixed $\kappa$. To make use of the delta function, it is convenient to change variable to integrate over $\Delta_\vec{q}^2$ instead of $q^2. $ Using Eq. \ref{q vs kappa} and \ref{eps expansion} we obtain
\be \label{q vs kappa 2}
q^2 \approx \kappa^2 \eta^2/\Delta_q^2.
\ee
After inserting the  term $d(q^2)/d\Delta_q^2$ in Eq. \ref{DOS int1}  we obtain
\be \label{dos int2}
\rho(\omega)= \frac{\eta^2}{\omega_L^2}\frac{\omega}{2\pi^2}\int_{0}^{\Lambda_q} d\kappa\int_{0}^{\Lambda_q} d\Delta_q^2 \,\delta(\Delta_{\omega}^2-\Delta_\vec{q}^2) \frac{\kappa^2 \eta^2}{\Delta_q^4}  e^{-2qz} e^{-\Delta_q^2/\eta^2}
\ee
Note that instead of the cut-off in the q integral, we have introduce an exponential function $e^{-\Delta_q^2/\eta^2}$ which serves the same purpose to limit the region of integration to where the approximation that $|\tilde{\epsilon}|<1$ is valid. The upper integration can be extended to infinity and we make use of the delta function to obtain
\be \label{dos int3}
\rho(\omega)= \frac{\eta^2}{\omega_L^2}\frac{\omega \eta^2}{2\pi^2 \Delta_{\omega}^4}\int_{0}^{\Lambda_q} d\kappa \, \kappa^2   e^{-2(\frac{\kappa\eta}{\Delta_{\omega}})z} e^{-\Delta_{\omega}^2/\eta^2}
\ee
The last factor $e^{-\Delta_{\omega}^2/\eta^2}$ can be set to unity under the expansion condition $\Delta_{\omega}^2/\eta^2 << 1$. The integral is easily done by rescaling $\kappa$ leading to
\be \label{DOS1}
\rho(\omega)= \frac{2\eta \omega }{2\pi^2 \omega_L^2\Delta_{\omega}}\frac{1}{(2z)^3}
\ee
Using the vacuum DOS $\rho_{vac}=\frac{1}{3\pi^2}\frac{\omega^2}{c^3}$ and expanding to leading order in $\frac{\omega_L-\omega}{\eta}$ we obtain
\be \label{DoS2}
\frac{\rho(\omega)}{\rho_{vac}(\omega)}=\frac{\frac{2}{3}\eta}{\sqrt{\omega_L^2-\omega^2}}(\frac{\lambda}{4\pi z})^3
\ee
where $\lambda=2\pi c/\omega$ is the wavelength in vacuum. Note the $1/z^3$ dependence and the square root divergence as $\omega$ approaches $\omega_L$.

Next we consider the case with $\omega$ slightly larger than $\omega_T$ and proceed with the expansion in a similar way. We define
\be\label{Delta omega T}
\Delta_{\omega,T}^2 = \omega^2-\omega_T^2.
\ee
We assume the condition  $|\tilde{\epsilon}|>>1$, in which case $\tilde{\epsilon}$ is approximated as
\be\label{eps expansionT}
\tilde{\epsilon}(\omega)=-\eta^2/\Delta_{\omega,T}^2
\ee
 and $\tilde{\epsilon}$  diverges as $-1/(\omega-\omega_T)$ for $\omega$ just larger  than $\omega_T$. We have 
 \be \label{qT vs kappa 2}
q^2 \approx (\kappa^2/ \eta^2)\Delta_{q,T}^2.
\ee
where
\be\label{deltaT q}
\Delta_{\vec{q},T}^2=\omega_{\vec{q}\kappa}^2  -\omega_T^2.
\ee
The equation for $\rho(\omega)$ is similar to Eq.\ref{DOS int1} 
but the normalization factor  now depends on $\vq$ and $\kappa$ via the mode frequency (see appendix 2 ) and is very small near $\omega_T$. It is given by $f_{\vq\kappa}^2/2=\Delta_{q,T}^4/(\eta^2\omega_T^2)$ and it must be left inside the $q$ integration. After the change of integration variable, we  find 
\be \label{dosT int2}
\rho(\omega)= \frac{\omega}{2\pi^2\eta^2\omega_T^2}\int_{0}^{\Lambda_q} d\kappa\int_{0}^{\Lambda_q} d\Delta_{q,T}^2 \,\delta(\Delta_{\omega,T}^2-\Delta_{q,T}^2) \frac{\kappa^2\Delta_q^4 }{\eta^2}  e^{-2qz} 
\ee
After using Eq.\ref{qT vs kappa 2} for $q$ in the exponential and using the delta function, we obtain
\be \label{DOST3}
\rho(\omega)= \frac{\omega }{2\pi^2 \eta^4 \omega_T^2}\int_{0}^{\Lambda_q} d\kappa \, \kappa^2 \Delta_{\omega}^4   e^{-2(\frac{\kappa\Delta_{\omega}}{\eta})z} e^{-\Delta_{\omega}^2/\eta^2}
\ee
the $\kappa$ integral is done by scaling, resulting in
\be \label{DOST1}
\rho(\omega)= \frac{2\omega }{2\pi^2 \omega_T^2 }\sqrt{\frac{\omega^2-\omega_T^2}{\eta^2}}\frac{1}{(2z)^3}
\ee
and \be \label{DoST2}
\frac{\rho(\omega)}{\rho_{vac}(\omega)}=2\sqrt{\frac{\omega^2-\omega_T^2}{\eta^2}}(\frac{\lambda}{4\pi z})^3
\ee
which is valid for $\omega$ slightly larger than $\omega_T$.

The photon density of states can be computed for type 1 HM in a similar way, using the expression for mode function given in Eq. \ref{f kappa 1} and will not be repeated here.

\

\section{ Normalization factor.
}
\label{appendix: normalization}
We begin by deriving Eq.\ref{wf main}. We use the expressions given by Ashida et al.\cite{ashida2021cavity} where they used a geometry of a slab with thickness $d'$ with a thin gap of thickness $\delta$ in the middle. We assume that the physical effect at a distance z on top of a slab of thickness $d$ is half of the effect for a position in the middle of the gap in the  geometry of Ashida et al. with $\delta=2z$ and $d'=2d$. In their notation, we write the wavefunction of mode $n$ in the middle of the gap as $f_{\vq,n}=f_{\vq \kappa_n}/\sqrt{d'}$ where we explicitly display the thickness dependence. From their Eq.S11, we have
\be \label{f''}
(f''_{\vq,n})^2= f_{\vq,n}^2 \cos^2(\kappa_{\vq n}z)/\cosh^2(\nu_{\vq n}z) \approx 4 f_{\vq,n}^2 \cos^2(\kappa_{\vq n}z) e^{-2\nu_{\vq n}z}
\ee
In the second step we expanded the $cosh$ for large values of its argument.
Now use $f_{\vq,n}=f_{\vq \kappa_n}/\sqrt{d'}$ and $d'=2d$ we convert the factor 4 in Eq.\ref{f''} to 2 before taking the continuum large $d$ limit. After dividing by 2 to account for having a slab on one side only, we obtain Eq,\eqref{wf}. Note that the sum on n in Ashida et al. is over positive integers only. This accounts for the integration over positive $\kappa$ after we take the continuum limit.

In the remainder of the paper,we take $d$ goes to infinity. 
Using the expressions in Ashida et al. \cite{ashida2021cavity} supp material we specialize to the case of a slab of thickness d which we send to infinity, converting the sum over discrete modes to an integral over $\kappa$ along the $z$ direction. In that case that the wavefunction that leaks outside is much smaller than that inside the hBN slab and can be ignore for the normalization calculation.  We first consider the type 2 transverse HM  at 1370 $cm^{-1}$ and set the dielectric constant at the longitudinal mode to be $\epsilon_z= \epsilon_{\infty}$  The eigen mode for the electric field inside the slab is given for $\vq$ along $x$  by 
\be   \label{u}
\vec{u}_{\vq\kappa} =f_{\vq\kappa}[\cos(\kappa z) \vec{e}_x-\frac{i \tilde{\epsilon q}}{\kappa} \sin(\kappa z) \vec{e}_z] e^{iqx}
\ee
where $\vec{e}_x,\vec{e}_z$ are unit vectors along $x,z$. The second term is obtained from the first term by the condition $\nabla \cdot\vec{D}=0$. The corresponding phonon mode function is given by
\be \label{u theta}
\vec{u}_{\vq\kappa}^\theta =f_{\vq\kappa}[\frac{\eta \omega_T}{\omega_T^2-\omega_{\vec{q}\kappa}^2}\cos(\kappa z) \vec{e}_x -\frac{i \tilde{\epsilon q}}{\kappa}A_\theta \sin(\kappa z) \vec{e}_z] e^{iqx}
\ee
where $A_\theta=\frac{\eta_z\omega_z}{\omega_z^2-\omega_{\vec{q}\kappa}^2}$ is the property of the longitudinal mode and is small because $\eta_z/\omega_z$ is small. The second term in Eq.\ref{u theta} is therefor small compared with the second term in Eq.\ref{u} and will be neglected in the calculation of the normalization factor below, which imposes the condition 
\be
\int_{-d/2}^{d/2} dz [|\vec{u}_{\vq\kappa}(z)|^2+|\vec{u}_{\vq\kappa}^\theta (z)|^2] =1
\ee
This result in the normalized factor
\be \label{f_qk}
\frac{f_{\vec{q}\kappa}^2}{2}[1+(\frac{\tilde{\epsilon}(\omega_{\vec{q}\kappa}) q}{\kappa})^2 + (\frac{\eta \omega_T}{\omega_T^2-\omega_{\vec{q}\kappa}^2})^2]=1
\ee
The second term in Eq.\ref{f_qk} can be written as $\tilde{\epsilon}(\omega_{\vec{q}\kappa})$ using Eq.\ref{q vs kappa}.
We consider two limits. As $\omega_{\vec{q}\kappa}$ approaches $\omega_L$ this term in vanishes as $\tilde{\epsilon}(\omega_{\vec{q}\kappa})$ goes to zero and becomes negligible. Then we find
\be \label{f near L}
\frac{f_{\vq\kappa}^2}{2}\approx\frac{1}{1+(\omega_T/\eta)^2}=(\frac{\eta}{\omega_L})^2, \quad\omega_{\vec{q}\kappa}\rightarrow\omega_L,
\ee

Next we consider the case $\omega$ near $\omega_T$ where $\tilde{\epsilon}(\omega_{\vec{q}\kappa})$ diverges. The third term in Eq.\ref{f_qk} which came from the z component of the wavefunction $\vec{u}^\phi$ strongly diverges as $1/(\omega^2-\omega_T^2) ^2\propto \tilde{\epsilon}^2$ 
which dominate the second term which is proportional to $\tilde{\epsilon}$. Keeping the most divergent term  gives rise to a nomalization factor that vanishes as $f_{\vq\kappa}^2/2=\Delta_{\omega,T}^4/\eta^2\omega_T^2$. 
Using Eq.
\ref{q vs kappa} this can also be written as 
\be \label{f near T}
\begin{split}
f_{\vq\kappa}^2/2&\approx \eta^2/(\omega_T^2\tilde{\epsilon}^2)\\&= \frac{\eta^2}{\omega_T^2}(\frac{q^2}{\kappa^2})^2, \quad \omega_{\vec{q}\kappa}\rightarrow\omega_T
\end{split}
\ee
Is is useful to write the following interpolation formula which is approximately valid for all of $\vq,\kappa$ space.
\be  \label{interpolate}
f_{\vq\kappa}^2/2= \frac{\eta^2/\omega_L^2}{1+(\omega_T/\omega_L)^2(\kappa^2/q^2)^2}
\ee
In computing the self energy we often encounter the integral
\be \label{I kappa}
I_{\kappa}=\int_0^\infty d\kappa \frac{f_{\vec{q}\kappa}^2}{2}.
\ee
Using the interpolation form the integral is easily evaluated by scaling the variable $\kappa$ to $\kappa/q$ leading to
\be \label{I kappa 1}
I_{\kappa}=\frac{\pi}{2\sqrt{2}} \frac{\eta^2}{\omega_T \omega_L}\sqrt{\frac{\omega_T}{\omega_L}}q
\ee
where we have used $\int_0^\infty dx \, 1/(1+x^4)=\pi/(2\sqrt{2})$.

We can repeat the steps to find the normalization factor $f_{1,\vq\kappa}^2/2$ near a type 1 HM. The limiting forms near $\omega_{1L}$ and $\omega_{1T}$ are different and we find the following interpolation formula that is approximately valid for all $\kappa,\vec{q}$:
\be 
\label{f kappa 1}
f_{1,\vec{q}\kappa}^2/2= \frac{\eta_1^2/\omega_{1L}^2}{\kappa^2/q^2+(\omega_{1T}/\omega_{1L})^2q^2/\kappa^2}
\ee
We introduce the an integral $I_{1\kappa}$ similar to Eq.\ref{I kappa} and again do the integral by scalling. This time the contribution comes mainly from the region near the dashed line in Fig.\ref{fig:fig.1.jpg}, where $\kappa \approx q$ and a different integral $\int_0^\infty dx \, x^2/(1+x^2)=\pi/(2\sqrt{2})$ is involved. Interestingly the final result for the integral $I_{1 \kappa}$ is the exactly the same as Eq.\ref{I kappa 1}, once the "1" subscript is entered everywhere.

The normalization can include the effect of a finite slab thickness in the following way. The wavefunction extends as $e^{-qz}$ for a distance $z$ outside the surface. For the purpose of calculating the integral of the absolute value of the wavefunction over all space , this effectively extend the thickness of the slab from $d $ to $d+1/2q$. Therefore the normalization factor must be multiplied by the factor $1/(1+1/2qd)$ to give the finite $d$ correction. This factor can easily be inserted into the q integral in the equations for $\rho(\omega,z)$ and while there is a change in the overall factor which is a function of $z/d$, it does not change the $\omega$ dependence of the result.

\section{Comparison of Weyl gauge and Coulomb gauge.}
\label{appendix: weyl gauge}
In this appendix we will first reproduce the earlier Weyl-gauge calculation, and then show how this calculation is fixed under Weyl gauge by accounting for gauge-dependent electron operator.

Under Weyl gauge, $\vec{E}=-\p_t\vec{A}$ when $\phi=0$, the mode expansion Eq.~\eqref{eq:Eparallel-operator} fixes the vector potential at the metallic plane,
\begin{align}
\vec{A}(\vec{r},t)=&\frac{1}{\sqrt d}\sum_{\kappa>0}\int\!\frac{d^2q}{(2\pi)^2}
\hat{\vq}\,\frac{1}{\omega_{\vq\kappa}}\sqrt{\frac{\omega_{\vq\kappa}}{2\epsilon_\infty}}\,
f(\vq,\kappa,z)
\nonumber\\
&\times\left(a_{\vq\kappa}e^{i(\vq\cdot\vec{r}-\omega_{q\kappa} t)}
+a^\dagger_{\vq\kappa}e^{-i(\vq\cdot\vec{r}-\omega_{q\kappa} t)}\right),
\label{eq:A-operator}
\end{align}
whose amplitude is related to the Coulomb-gauge vertex Eq.~\eqref{eq:coulomb-vertex} by $eA_{\vq\kappa}/q=g_{\vq\kappa}/\omega_{\vq\kappa}$. Because each HM mode is polarized along its own $\hat{\vq}$, this $\vec{A}$ is curl free and is a pure gradient,
\begin{align}
\vec{A}(\vec{r},t)=\bm{\nabla}\chi(\vec{r},t),\quad
&\chi(\vec{r},t)=\frac{1}{\sqrt d}\sum_{\kappa>0}\int\!\frac{d^2q}{(2\pi)^2}
\frac{1}{i}\,\frac{g_{\vq\kappa}}{e\,\omega_{\vq\kappa}}
\nonumber\\
&\times\left(a_{\vq\kappa}e^{i(\vq\cdot\vec{r}-\omega_{q\kappa} t) }
-a^\dagger_{\vq\kappa}e^{-i(\vq\cdot\vec{r}-\omega_{q\kappa} t)}\right),
\label{eq:chi}
\end{align}
a fact we shall use twice below. The coupling is now $e\int d^2r\,\vec{j}\cdot\vec{A}$. Only the longitudinal projection of the current enters, and the continuity equation $\vq\cdot\vec{j}_{\vq}=[H_e,\rho_{\vq}]$ converts it into an energy difference, so that
\begin{align}
H_{jA}&=\frac{1}{\sqrt d}\sum_{\vq,\kappa>0}\sum_{\vk}
g^W_{\vk\vq\kappa}\,c^\dagger_{\vk+\vq}c_{\vk}
\left(a_{\vq\kappa}+a^\dagger_{-\vq\kappa}\right),
\nonumber\\
g^W_{\vk\vq\kappa}&=g_{\vq\kappa}\,
\frac{\xi_{\vk+\vq}-\xi_{\vk}}{\omega_{\vq\kappa}} .
\label{eq:gW}
\end{align}
Here and below we abbreviate $\Delta\xi\equiv\xi_{\vk+\vq}-\xi_{\vk}$. The vertex now carries the electronic energy difference in place of the mode frequency; the two coincide only on shell, $|\Delta\xi|=\omega_{\vq\kappa}$. (The $A^2$ term gives a frequency independent constant which is absorbed into the chemical potential, as before.)

Repeating the one-loop calculation of the previous section with $g_{\vq\kappa}\to g^W_{\vk\vq\kappa}$ in Eq.~\eqref{eq:retarded-self-energy}, and differentiating as in Eq.~\eqref{eq:lambda-def}, gives
\begin{align}
\lambda^W_0=&\frac{1}{d}\sum_{\kappa>0}\int\!\frac{d^2q}{(2\pi)^2}
|g_{\vq\kappa}|^2\frac{1}{\left(|\xi_{\vk_F+\vq}|+\omega_{\vq\kappa}\right)^2}\,
\frac{\xi^2_{\vk_F+\vq}}{\omega^2_{\vq\kappa}},
\label{eq:lambda0_W}
\end{align}
which differs from Eq.~\eqref{eq:lambda-q} by the factor $\xi^2_{\vk_F+\vq}/\omega^2_{\vq\kappa}$ under the integral. Reducing it exactly as before, with $x=qz$ and $\xi_{\vk_F+\vq}\simeq v_Fq\cos\theta$ so that $\xi/\bar\omega=\alpha x\cos\theta$,
\begin{align}
\lambda^W_0&=C\alpha
\int_0^\infty\frac{dx}{\cosh^2x}
\int_0^{2\pi}\frac{d\theta}{2\pi}
\frac{\left(\alpha x\cos\theta\right)^2}
{\left(1+\alpha x|\cos\theta|\right)^2}
\nonumber\\
&\simeq
\begin{cases}
C\alpha^3, & \alpha\ll1,\\[1mm]
C\alpha, & \alpha\gg1 .
\end{cases}
\label{eq:lambda0_W_1}
\end{align}
Clearly this naive Weyl-gauge result disagrees with the Coulomb-gauge result in Eq.~\eqref{eq:lambda-asymptotics}

The step that fails comes before Eq.~\eqref{eq:lambda0_W}.  Both the self energy Eq.~\eqref{eq:retarded-self-energy} and the residue Eq.~\eqref{eq:Z-fermion} refer to the electron Green function, and carrying them over to the Weyl gauge means we were essentially using the following form of green's function in the Weyl-gauge derivation:
\begin{equation}
G(\vk,t)=-i\big\langle T\, c_{\vk}(t)\,c^\dagger_{\vk}(0)\big\rangle .
\label{eq:Gassumed}
\end{equation}
where $c_{\vk}(t)=e^{iHt}c_{\vk}e^{-iHt}$ is the Heisenberg-picture electron operator in the Weyl gauge, whereas $H$ is the full Hamiltonian. However, this is wrong because the Green's function we should extract Z factor from is the gauge-invariant one, which should be written as 
\begin{equation}
G(\vk,t)=-i\big\langle T\,\tilde c_{\vk}(t)\,\tilde c^\dagger_{\vk}(0)\big\rangle ,
\qquad
\tilde c(\vec{r},t)=e^{ie\chi(\vec{r},t)}c(\vec{r},t) ,
\label{eq:Gcorrect}
\end{equation}
where $c(\vec{r},t)=e^{iHt}c(\vec{r})e^{-iHt}$ Heisenberg-picture is the electron field of the gauge we are working in (Weyl gauge here), whereas $\tilde c(\vec{r},t)$ is the gauge-invariant (physical) electron operator. Here $\chi$ is defined so that $\bm{\nabla}\chi$ is the longitudinal part of the vector potential of that gauge, whereas $c_{\vk}(t)$ and $\tilde c_{\vk}(t)$ are the Fourier transforms of $\tilde c(\vec{r},t)$ and $\tilde c(\vec{r},t)$. 

In the Coulomb gauge the vector potential is transverse, so $\chi=0$, $\tilde c=c$, and Eq.~\eqref{eq:Gassumed} is the Green function of Eq.~\eqref{eq:Gcorrect}. Therefore, the calculation of Sec.~III stands as it is.  In the Weyl gauge $\chi$ is the operator of Eq.~\eqref{eq:chi}. The gauge-invariant electron operator in Schrodinger picture $\tilde{c}(\vec{r})$ can be obtained by taking $t=0$ in the expression of Heisenberg-picture operator $\tilde{c}(\vec r,t)$ and expanding it to first order: $\tilde c(\vec{r})=\tilde{c}(\vec r,t=0)=(1+ie\chi(\vec r,0))c(\vec{r})+O(e^2)$.
Using $\chi$ from Eq.~\eqref{eq:chi}, we find
\begin{align}
\tilde c_{\vk}\approx &\,c_{\vk}
+\frac{1}{\sqrt d}\sum_{\vq,\kappa>0}\frac{g_{\vq\kappa}}{\omega_{\vq\kappa}}\left(a_{\vq\kappa}c_{\vk-\vq}e^{i\vq\cdot\vec{r}}-a^\dagger_{\vq\kappa}c_{\vk+\vq}e^{-i\vq\cdot\vec{r}}\right).
\label{eq:ctilde}
\end{align}


We now evaluate the residue of Eq.~\eqref{eq:Gcorrect} in the Weyl gauge, taking $\vk$ on the Fermi surface.  Inserting eigenstates of the full Hamiltonian, $H|m\rangle=E_m|m\rangle$ and $H|G\rangle=E_G|G\rangle$ for the ground state, the removal part of Eq.~\eqref{eq:Gcorrect} at $t>0$ is
\begin{equation}
\langle G|\tilde c^\dagger_{\vec k_F}\tilde c_{\vec k_F}(t)|G\rangle
=\sum_m e^{i(E_m-E_G)t}\big|\langle m|\tilde c_{\vec k_F}|G\rangle\big|^2 ,
\label{eq:insert}
\end{equation}
In the Lehmann representation, the spectral function of Eq.~\eqref{eq:Gcorrect} is
\begin{align}
A(\vk=\vec k_F,\omega)=&\sum_m\big|\langle m|\tilde c_{\vec k_F}|G\rangle\big|^2
\,\delta\!\left(\omega-E_G+E_m\right)
\nonumber\\
&+\sum_{m'}\big|\langle m'|\tilde c^\dagger_{\vec k_F}|G\rangle\big|^2
\,\delta\!\left(\omega-E_{m'}+E_G\right),
\label{eq:Lehmann}
\end{align}
where $|G\rangle$ is the ground state, 
the summation over $m$ and $m'$ runs over all exact eigenstates with one electron removed ($|m\rangle$) and one added ($|m'\rangle$).

The phases in $\tilde c$ cancel at coincident points, so $\{\tilde c_{\vk},\tilde c^\dagger_{\vk}\}=1$ exactly as for $c_{\vk}$, the total weight in Eq.~\eqref{eq:Lehmann} is unity, and
\begin{equation}
1-Z=\sum_{m\neq{\rm qp}}\big|{\cal M}_{m}\big|^2
+\sum_{m'\neq{\rm qp}}\big|{\cal M}_{m'}\big|^2 ,
\label{eq:oneminusZ}
\end{equation}
where ${\cal M}_{m}=\langle m|\tilde c_{\vec k_F}|G\rangle$ and ${\cal M}_{m'}=\langle m'|\tilde c^\dagger_{\vec k_F}|G\rangle$ are the amplitudes for reaching the excited states of the two sectors.  Here $\sum_{m\neq{\rm qp}}$ means the summation excludes the whole Hilbert space spanned by quasiparticle  basis $\lbrace  c_{\vec k_F}|{\rm FS};0\rangle  \rbrace$.

Next, we express the states in ${\cal M}_m$ in terms of HM-electron coupling. These states can be considered as perturbed by the $j\cdot A$ coupling from eigenstates of noninteracting HM and electron Hamiltonian $|m^{(0)}\rangle$ and $|G^{(0)}\rangle=|{\rm FS};0\rangle$, i.e.
\begin{align}
|m\rangle&=|m^{(0)}\rangle+|m^{(1)}\rangle ,
&
|m^{(1)}\rangle&=\sum_{n}\frac{\langle n^{(0)}|H_{jA}|m^{(0)}\rangle}{E^{(0)}_m-E^{(0)}_n}|n^{(0)}\rangle ,
\nonumber\\
&&|m^{(0)}\rangle&=a^\dagger_{\vq\kappa}c_{\vec{k_F}+\vq}|{\rm FS};0\rangle ,
\nonumber\\
|G\rangle&=|{\rm FS};0\rangle+|G^{(1)}\rangle ,
&
|G^{(1)}\rangle&=\sum_{n}\frac{\langle n^{(0)}|H_{jA}|{\rm FS};0\rangle}{E_{\rm FS}-E^{(0)}_n}|n^{(0)}\rangle ,
\label{eq:states}
\end{align}
with $\xi_{\vec k_F+\vq}<0$ in the removed electron sector, $\xi_{\vec k_F +\vq}>0$ in the added electron sector.
The sums over $n$ running over the unperturbed eigenstates. From now on we relabel the eigenstates using their unperturbed state to explicitly show its components: $|m\rangle \rightarrow |\vec q \kappa \rangle$ .
 Now we can calculate the amplitude $\mathcal{M}_m$ which is now relabeled as $\mathcal{M}_{\vq\kappa}=\langle \vq\kappa|\tilde c_{\vec k_F}|G\rangle$. The amplitude then has three pieces at $O(e^2)$: ${\cal M}_{\vq\kappa}={\cal M}_{\vq\kappa}^{(a)}+{\cal M}_{\vq\kappa}^{(b)}+{\cal M}_{\vq\kappa}^{(c)}$, with
\begin{align}
{\cal M}_{\vq\kappa}^{(a)}&=\langle m^{(0)}|\,c_{\vec k_F}\,|G^{(1)}\rangle ,
\nonumber\\
{\cal M}_{\vq\kappa}^{(b)}&=\langle m^{(1)}|\,c_{\vec k_F}\,|{\rm FS};0\rangle ,
\nonumber\\
{\cal M}_{\vq\kappa}^{(c)}&=-\frac{1}{\sqrt d}\frac{g_{\vq\kappa}}{\omega_{\vq\kappa}}\,
\langle m^{(0)}|\,a^\dagger_{\vq\kappa}c_{\vec k_F+\vq}\,|{\rm FS};0\rangle .
\label{eq:three}
\end{align}
The first term ${\cal M}_{\vq\kappa}^{(a)}$ vanishes due to mismatch of the electron number. Using $E^{(0)}_m-E^{(0)}_n=-(\Delta\xi-\omega_{\vq\kappa})$, we find ${\cal M}_{\vq\kappa}^{(b)}$ is given by
 \begin{equation}
{\cal M}_{\vq\kappa}^{(b)}=\frac{1}{\sqrt d}\frac{g^W_{\vq\kappa}}{\Delta\xi-\omega_{\vq\kappa}} ,
\qquad
\sum_{\vq\kappa}\big|{\cal M}_{\vq\kappa}^{(b)}\big|^2=\lambda^W_0 ,
\label{eq:Mb}
\end{equation}
so this piece by itself is the amplitude we obtained in the naive Weyl-gauge derivation Eq.~\eqref{eq:lambda0_W_1}. Now we see what we missed: the ${\cal M}_{\vq\kappa,\vec k}^{(c)}$ term.
In ${\cal M}_{\vq\kappa}^{(c)}$ the operator $a^\dagger_{\vq\kappa}c_{\vec k_F+\vq}$ produces $|m^{(0)}\rangle$ out of $|{\rm FS};0\rangle$, so the matrix element is unity and
\begin{equation}
{\cal M}_{\vq\kappa}^{(c)}=-\frac{1}{\sqrt d}\frac{g_{\vq\kappa}}{\omega_{\vq\kappa}} .
\label{eq:Mc}
\end{equation}
Adding the three and using $g^W_{ \vq\kappa}=g_{\vq\kappa}\Delta\xi/\omega_{\vq\kappa}$,
\begin{align}
{\cal M}_{\vq\kappa,\vec k}
&=\frac{1}{\sqrt d}\frac{g_{\vq\kappa}}{\omega_{\vq\kappa}}\,
\frac{\Delta\xi-\left(\Delta\xi-\omega_{\vq\kappa}\right)}
{\Delta\xi-\omega_{\vq\kappa}}
=\frac{1}{\sqrt d}\frac{g_{\vq\kappa}}{\Delta\xi-\omega_{\vq\kappa}} .
\label{eq:Mtot}
\end{align}
This recovers the correct electron-HM vertex we found in Coulomb-gauge derivation.
The phase cancels the electronic energy difference in the numerator and puts the mode frequency in its place.  The added electron sector runs in parallel, with the energy denominator $\Delta\xi+\omega_{\vq\kappa}$ and $\xi_{\vk+\vq}>0$.  Summing the two sectors, which together cover the whole $\vq$ integration, and using $|\Delta\xi\mp\omega_{\vq\kappa}|=|\xi_{\vk+\vq}|+\omega_{\vq\kappa}$,
\begin{align}
1-Z&=\sum_{\kappa>0}\int\!\frac{d^2q}{(2\pi)^2} |{\cal M}_{\vq\kappa}|^2
\nonumber
\\
&\simeq\frac{1}{d}\sum_{\kappa>0}\int\!\frac{d^2q}{(2\pi)^2}
\frac{|g_{\vq\kappa}|^2}{\left(|\xi_{\vk+\vq}|+\omega_{\vq\kappa}\right)^2}
=\lambda_0 ,
\label{eq:Zfinal}
\end{align}
Therefore, the Weyl gauge calculation, done with the gauge-invariant Green function, recovers the results of Sec.~\ref{sec:Coulomb gauge}, Eqs.~\eqref{eq:lambda-asymptotics} and \eqref{eq:Z-fermion}, which are the correct results.


\newpage
\end{widetext}

\bibliography{ref}
\end{document}